\documentclass[a4paper,fleqn]{arxiv}

\usepackage{graphicx}
\usepackage{epstopdf, epsfig}
\usepackage{booktabs}
\usepackage{multirow}
\usepackage{array}
\usepackage{calc}
\usepackage{microtype}
\usepackage{lscape}
\usepackage{esint,upgreek}
\usepackage{amsmath,amsfonts}
\usepackage{bm}
\usepackage{afterpage}
\usepackage{blindtext}
\usepackage{multicol}

\usepackage{framed} % Framing content
\usepackage{multicol} % Multiple columns environment
\usepackage{nomencl} % Nomenclature package
\makenomenclature
\renewcommand*\nompreamble{\begin{multicols}{2}}
\renewcommand*\nompostamble{\end{multicols}}

\usepackage{algorithm2e}
\RestyleAlgo{ruled}
\usepackage[normalem]{ulem}
\usepackage{color}
\usepackage{mylines}
\usepackage{myplots}
\newcommand{\sy}[2]{\mbox{(\kern-.25em\SymbolRGB[solid]{#1}{.8pt}{#2}{5pt}\kern-.25em)}}
\newcommand{\lsy}[3]{\mbox{(\kern-.1em\lineSymbolRGB{#1}{#2}{2pt}{#3}{4pt}\kern-.45em)}}
\newcommand{\lcap}[2]{~\,{\kern-1em\protect\mylcap{#1}{#2}}}

\definecolor{blue}{rgb}{0,0,1}
\definecolor{red}{rgb}{1,0,0}
\definecolor{black}{rgb}{0,0,0}
\definecolor{grey}{rgb}{0.775,0.775,0.775}
\definecolor{white}{rgb}{1,1,1}

\definecolor{BlueTBL1}{rgb}{0.41961,0.682353,0.839216}
\definecolor{BlueTBL2}{rgb}{0.12941,0.443137,0.709804}
\definecolor{green_laser}{RGB}{0,208,0}
\definecolor{green_lgac}{rgb}{0.1068,0.5190,0.2488}
\definecolor{red_lgac}{rgb}{0.7630,0.0863,0.1068}
\definecolor{blue_lgac}{rgb}{0.1076,0.4153,0.6880}
\definecolor{red_pod_1}{rgb}{0.9467,0.2682,0.1961}
\definecolor{red_pod_2}{rgb}{0.7365,0.0800,0.1012}
\definecolor{red_pod_3}{rgb}{0.4039,0,0.0510}

\newcommand{\Ren}{\mathrm{Re}}
\newcommand{\Nu}{\mathrm{Nu}}

\usepackage[numbers]{natbib}

\def\tsc#1{\csdef{#1}{\textsc{\lowercase{#1}}\xspace}}
\tsc{WGM}
\tsc{QE}
\begin{document}
\let\WriteBookmarks\relax
\def\floatpagepagefraction{1}
\def\textpagefraction{.001}

% Short title
\shorttitle{Unsteady Thermal and Flow Structures of an Impinging Sweeping Jet}
% Short author
\shortauthors{R. Castellanos \textit{et al.}}  

% Main title of the paper
\title [mode = title]{Unsteady Thermal and Flow Structures of an Impinging Sweeping Jet}

%--------------- Rodri
\author[1]{Rodrigo Castellanos}[
    orcid=0000-0002-7789-5725]
\ead{rcastell@ing.uc3m.es}
\credit{Conceptualization, Methodology, Software, Validation, Formal analysis, Investigation, Data Curation, Writing - Original Draft, Writing - Review \& Editing, Visualization, Supervision, Project administration}
\affiliation[1]{organization={Department of Aerospace Engineering, Universidad Carlos III de Madrid},
            city={Legan\'es},
            postcode={28911}, 
            state={Madrid},
            country={Spain}}
%--------------- Adri
\author[1]{Adri\'an Mart\'in-Perrino}[
    orcid=0009-0000-2171-7957]
\ead{adrian.m.perrino@alumnos.uc3m.es}
\credit{Methodology, Software, Validation, Formal analysis, Investigation, Data Curation, Writing - Original Draft, Writing - Review \& Editing, Visualization}
%--------------- Elena
\author[2]{Elena L\'opez-N\'u\~nez}[
    orcid=0000-0002-9076-2898]
\ead{elena.lopez.nunez@upm.es}
\credit{Investigation, Supervision, Writing - Original Draft, Writing - Review \& Editing}
\affiliation[2]{organization={Instituto Universitario de Microgravedad “Ignacio Da Riva” (IDR/UPM), ETSI Aeronáutica y Del Espacio, Universidad Politécnica de Madrid},
            address={Pza. Del Cardenal Cisneros 3},
            city={Madrid},
            postcode={28040}, 
            state={Madrid},
            country={Spain}}
%--------------- Carlos
\author[3,1]{Carlos {Sanmiguel Vila}}[
    orcid=0000-0003-0331-2854]
\ead{csanmigu@ing.uc3m.es}
\cormark[1] % Corresponding author indication
\credit{Conceptualization, Methodology, Software, Validation, Formal analysis, Investigation, Data Curation, Resources, Writing - Original Draft, Writing - Review \& Editing, Supervision, Project administration, Funding acquisition}
\affiliation[3]{organization={Aerial Platforms Department, Spanish National Institute for Aerospace Technology (INTA)},
            city={San Mart\'in de la Vega},
            postcode={28330}, 
            state={Madrid},
            country={Spain}}

% % Corresponding author text
\cortext[1]{Corresponding author}

\maketitle

%%%%%%%%%%%%%%%%%%%%%%%%%%% ABSTRACT %%%%%%%%%%%%%%%%%%%%%%%%%%%%%%%%
\noindent \textsc{Abstract}
\vspace{0.15cm}

\noindent Sweeping jets are increasingly employed in thermal management and flow control applications due to their inherent unsteadiness and ability to cover wide surface areas. This study investigates the unsteady heat transfer and flow structures of an impinging sweeping jet, with particular emphasis on the influence of Reynolds number ($\Ren$) and nozzle-to-plate spacing ($L/D$). A combined diagnostic approach, utilising non-synchronous high-speed infrared thermography and planar particle image velocimetry, independently measures wall heat flux distributions and flow field dynamics. Time-averaged Nusselt number maps reveal a progressive widening of the thermal footprint as the nozzle-to-plate distance increases, accompanied by a reduction in peak heat transfer intensity. In contrast, the fluctuating component exhibits a double-lobe structure that becomes more pronounced at higher $\Ren$. Velocity measurements confirm the presence of a bifurcated jet core and shear-driven unsteadiness, with fluctuating velocity components and Reynolds shear stress spatially correlated with zones of enhanced heat transfer fluctuations. A direct spatial comparison between velocity and heat transfer fluctuations demonstrates that wall-normal velocity unsteadiness dominates the thermal response at low nozzle-to-plate distances, while streamwise fluctuations become increasingly significant at larger spacings. Finally, the full-field heat transfer data are embedded into a reduced-order parametric space, where the complex system behaviour is effectively represented by a low-dimensional manifold governed by $\Ren$ and $L/D$. This manifold is used to perform an efficient surrogate model capable of predicting both mean and fluctuating Nusselt number maps for untested combinations of $\Ren$ and $L/D$ within the parametric space, serving as a tool for the preliminary sizing and design of heat transfer enhancement devices.

%%%%%%%%%%%%%%%%%%%%%%%%%%%%%%%%%%%%%%%%%%%%%%%%%%%%%%%%%%%%%%%%%%%%%%%%%%%%%%%%%%%%%%%%%%%%
\vspace{0.4cm}
\noindent \textsc{Keywords}
\vspace{-0.2cm}
\begin{multicols}{2}
\noindent Sweeping Jet \\ Impinging flows \\ Infrared thermography \\ Unsteady heat transfer \\ Particle Image Velocimetry \\ Manifold learning
\end{multicols}

%%%%%%%%%%%%%%%%%%%%%%%%%%%%%%%%%%%%%%%%%%%%%%%%%%%%%%%%%%%%%%%
%%%%%%%%%%%%%%%%%%%%%%%%%%% nomenclature %%%%%%%%%%%%%%%%%%%%%%
%%%%%%%%%%%%%%%%%%%%%%%%%%%%%%%%%%%%%%%%%%%%%%%%%%%%%%%%%%%%%%%
\begin{table*}[hbt]   
    \begin{framed}
    % Greek letters:
    \nomenclature[00]{$\alpha$}{Thermal diffusivity}
    \nomenclature[01]{$\beta_v$}{Volumetric thermal expansion coefficient}
    \nomenclature[02]{$\varepsilon$}{Emissivity}
    \nomenclature[03]{$\nu$}{Kinematic viscosity}
    \nomenclature[04]{$\rho$}{Density}
    \nomenclature[05]{$\sigma$}{Boltzmann constant} 
    \nomenclature[05]{$\sigma_o$}{Standard deviation of magnitude $o$} 
    % Latin letters:
    \nomenclature[06]{$A_\mathrm{foil}$}{Area of the heated-thin-foil sensor}
    \nomenclature[07]{$a_g$}{Acceleration of gravity}
    \nomenclature[08]{$D$}{Hydraulic diameter, throat width}
    \nomenclature[09]{$\mathrm{Bi}$}{Biot number}
    \nomenclature[10]{$\mathrm{Fo}$}{Fourier number}
    \nomenclature[10]{$f$}{Frequency of the fluidic oscillator}
    \nomenclature[11]{$f_\#$}{Focal ratio}
    \nomenclature[12]{$\mathrm{Gr}$}{Grashoff number}
    \nomenclature[13]{$H$}{Height of the nozzle exit}
    \nomenclature[14]{$h$}{Convective heat transfer coefficient}
    \nomenclature[20]{$s_i$}{Thickness of layer $i$}
    \nomenclature[21]{$c_i$}{Heat capacity of layer $i$}
    \nomenclature[35]{$I$}{Electric current}
    \nomenclature[42]{$k_i$}{Thermal conductivity of element $i$}
    \nomenclature[43]{$L$}{Nozzle-to-plate distance}
    \nomenclature[44]{$\dot{m}$}{Mass flow rate}
    \nomenclature[48]{$\Nu$}{Nusselt number}
    \nomenclature[49]{$\Nu'$}{Fluctuating Nusselt number}
    \nomenclature[53]{$q''_{j}$}{Heat flux by Joule effect}
    \nomenclature[54]{$q''_{k}$}{Heat flux by conduction}
    \nomenclature[55]{$q''_{r}$}{Heat flux by radiation}
    \nomenclature[53]{$q''_{u}$}{Unsteady heat flux term}
    \nomenclature[57]{$\Ren$}{Reynolds number}
    \nomenclature[61]{$\mathrm{St}$}{Strouhal number}
    \nomenclature[63]{$T_\mathrm{w}$}{Wall temperature}
    \nomenclature[64]{$T_\mathrm{aw}$}{Adiabatic wall temperature}
    \nomenclature[65]{$T_\infty$}{Freestream temperature}
    \nomenclature[70]{$t_i$}{Thickness of element $i$}
    \nomenclature[71]{$U,V$}{Wall-normal and wall-tangent velocity}
    \nomenclature[72]{$U_c$}{Characteristic bulk velocity of the jet}
    \nomenclature[74]{$u,v$}{Wall-normal and wall-tangent velocity fluctuations}
    \nomenclature[76]{$V_{DC}$}{Electric voltage}
    \nomenclature[77]{$x,y,z$}{Spanwise, wall-tangent, and wall-normal directions}
    % Operadores:
    \nomenclature[81]{$\widetilde{\bullet}$}{Normalise with the bulk velocity operator}
    \nomenclature[82]{$\overline{\hspace{0.5mm} \bullet \hspace{0.5mm}}$}{Time-average operator}
    \nomenclature[83]{$\langle{\bullet}\rangle$}{Spatial-average operator}    
    \printnomenclature
    \end{framed}
\end{table*}

%%%%%%%%%%%%%%%%%%%%%%%%%%%%%%%%%%%%%%%%%%%%%%%%%%%%%%%%%%%%%%%%%%%%%
%%%%%%%%%%%%%%%%%%%%%%%%%%% INTRODUCTION %%%%%%%%%%%%%%%%%%%%%%%%%%%%
%%%%%%%%%%%%%%%%%%%%%%%%%%%%%%%%%%%%%%%%%%%%%%%%%%%%%%%%%%%%%%%%%%%%%
%\the\linewidth
%\the\textwidth
\newpage
\section{Introduction}\label{s:intro}
% Motivation of Impinging jets
In many engineering applications, ranging from electronics cooling to aerospace and turbine blade thermal management, there is a growing demand for compact and highly efficient convective cooling solutions. Among these, impinging jets stand out as one of the most effective configurations for achieving high heat transfer rates, particularly when surface temperature uniformity and localised thermal management are critical~\citep{meola2009new}. Since the seminal study by \citet{gardon1962heat}, extensive research has been devoted to understanding the influence of various parameters and configurations, as summarised in several comprehensive reviews~\citep{viskanta1993heat,carlomagno2014thermo,ekkad2021modern}. In applications employing fixed nozzle geometries and air as the working fluid, two primary non-dimensional parameters govern the heat transfer performance: the jet Reynolds number ($\Ren$) and the nozzle-to-plate distance ($L/D$) \citep{carlomagno2014thermo}. The latter is particularly influential, as positioning the plate beyond the potential core length (typically 4--6 nozzle diameters) promotes the development of centreline turbulence, which can enhance stagnation-point heat transfer. However, optimal $L/D$ values often reflect a compromise between competing effects. In addition, the nozzle exit geometry plays a pivotal role by dictating the initial shear layer development. This, in turn, affects turbulence production, fluid entrainment, and jet spreading rates, thereby shaping the spatial distribution of heat transfer on the impingement surface \citep{ekkad2021modern}. As a result, different nozzle designs can lead to markedly distinct thermal performance characteristics.

% Active control vs passive in impinging jets
Beyond the optimisation of steady jet configurations, there is increasing interest in exploiting unsteady jets to enhance convective heat transfer and improve surface coverage uniformity. Jet pulsations or oscillations can be generated through various approaches, including active forcing, as in synthetic jets~\citep{greco2018effects,krishan2019synthetic,glezer2002synthetic}, or passive self-exciting mechanisms such as annular, swirling, and sweeping jets \citep{bisht2023jet}. While active devices offer precise control, they typically require external actuators and complex integration. In contrast, passive configurations induce oscillations through internal flow dynamics and geometric design, eliminating the need for moving parts. Examples include annular jets, which exploit vortex shedding~\citep[e.g.][]{terekhov2016experimental}, and swirling jets, where natural instabilities such as the precessing vortex core dominate the unsteady motion~\citep{ianiro2012heat}.

% Sweeping jets in particular
Among passive methods, sweeping jets generated by fluidic oscillators are particularly relevant for impinging applications. These devices convert a steady inflow into an oscillating jet through internal feedback mechanisms~\citep{ostermann2018properties,woszidlo2019fundamental}. Typically, the incoming flow enters an interaction chamber equipped with lateral feedback channels (see the schematic in \autoref{fig:setup}). The jet initially attaches to one sidewall via the Coand\u{a} effect, while part of the flow is redirected through a feedback channel to destabilise this attachment. This cyclical process forces the jet to switch sides periodically, producing a stable oscillation that propagates downstream. When discharged through a divergent nozzle, this oscillating flow attaches alternately to the nozzle sidewalls, resulting in a large-amplitude, lateral sweeping motion.
Compared to steady impinging jets, sweeping jets inherently introduce strong unsteadiness and impact a broader wall area, even at moderate Reynolds numbers, making them attractive for applications requiring enhanced mixing or improved thermal uniformity.

% Impinging SJ
Despite their potential advantages, sweeping jets have received comparatively limited attention in the context of impinging heat transfer, particularly when contrasted with conventional steady jets. Early experimental studies have nevertheless provided valuable insights into their time-averaged behaviour. For instance, \citet{kim2019measurement} demonstrated that sweeping jets can achieve higher mean convective heat transfer coefficients than conventional round jets, especially at short nozzle-to-plate distances ($L/D < 5$). This enhancement was attributed to intensified turbulent mixing generated by the sweeping motion. However, their study also revealed that the heat transfer benefit tends to diminish as $L/D$ increases. Velocity field measurements suggested that, when time-averaged, the jet resembled two oblique streams impinging near the lateral extremities of the sweep, reflecting the prolonged residence time of the jet attached to the nozzle sidewalls. This resulted in a strong spatial correlation between the time-averaged velocity patterns and the distribution of the time-averaged Nusselt number on the target surface. Complementing velocity and heat transfer studies, \citet{park2018flow} applied Proper Orthogonal Decomposition (POD) to correlate dominant flow structures with radial distributions of the Nusselt number. Their results linked the first POD mode to bi-stable sweeping behaviour, highlighting the influence of flow topology on surface heat transfer. However, their analysis was limited to time-averaged data and did not capture the fluctuating component of the heat transfer field, which is essential for understanding the unsteady dynamics of sweeping jet impingement, particularly because such fluctuations can induce significant thermal stresses due to rapid and localised temperature variations.

More recent experimental studies have further characterised this behaviour. \citet{gomes2024flow}, for example, showed that the advantage of sweeping jets in terms of heat transfer uniformity may persist at larger $L/D$ values under low-Reynolds-number conditions. Meanwhile, \citet{dangelo2024piv} conducted a systematic Particle Image Velocimetry (PIV) study to investigate the effects of internal oscillator geometry and $L/D$ on jet dynamics. Their results confirmed the significant influence of these parameters on the oscillation frequency, the presence of sweeping motion, and the rigidity of the jet near the impingement surface. Notably, they observed that sweeping coherence decreased with increasing $L/D$, which has direct implications for the resulting wall heat transfer distribution.

% Numerical & Complex Geom
Numerical simulations complemented these experimental findings, particularly for parametric investigations relevant to industrial applications such as turbine blade cooling. A series of RANS and URANS studies~\citep{hossain2021effects, joulaei2022parametric, joulaei2023numerical, khan2023comparison} reproduced key trends observed experimentally, including the reduction in average heat transfer at higher $L/D$ ratios. These works have also highlighted the sensitivity of the flow to geometric features, such as the divergent nozzle angle. For example, \citet{hossain2021effects} reported that wider nozzle exit angles tended to reduce the mean heat transfer, possibly due to flow separation, although they simultaneously contributed to more uniform surface temperature distributions. These geometric effects are particularly relevant in complex internal configurations, such as curved passages within turbine blades. Studies of leading-edge cooling models have demonstrated that sweeping jet arrays can increase the Nusselt number by 12-15\% compared to steady or non-periodic jets at equivalent mass flow rates~\citep{Ten2019turbo,khan2023comparison}. This improvement extends to other constrained geometries, including confined concave surfaces, where sweeping jets have achieved up to 13\% higher heat transfer than conventional circular jets, primarily by maintaining effective wall coverage and strong mixing despite the geometric complexity~\citep{tu2023flow}. Moreover, the application of sweeping jets has been extended to external film cooling configurations. Studies analysing the spatio-temporal distribution of the coolant film have shown that sweeping jets can provide enhanced cooling effectiveness and spatial uniformity compared to conventional approaches~\citep{zhou2022pof}. Subsequent investigations have demonstrated that further improvements can be achieved by modifying the nozzle exit shape, thereby extending the applicability of sweeping jets to demanding environments such as transonic turbine vanes~\citep{wang2024turbomach,wang2025ijts}.

%% New Sweeping jets
Recent research efforts have increasingly focused on optimising sweeping jet geometry to further enhance thermal performance. For example, \citet{tang2024adjoint} applied adjoint-based shape optimisation to redesign the internal flow path of the oscillator for jet impingement on concave surfaces. This optimization resulted in a more focused jet with a reduced oscillation angle, leading to a 11.6\% increase in the time- and surface-averaged Nusselt number compared to conventional oscillator configurations. Complementing this, recent studies have proposed modified passive oscillators with adjustable internal flow features to enable real-time tuning of jet characteristics~\citep{martinke2025characteristics}. Furthermore, hybrid passive-active designs with external actuation have been shown to enhance stagnation-point heat transfer at low $H/D$ ratios and improve temperature uniformity at larger distances~\citep{donofrio2024synthetic}, offering flexibility for complex thermal management scenarios. These developments, along with active control strategies based on master–slave actuator configurations~\citep{wen2020jet}, highlight a promising path toward real-time adaptive cooling technologies.

% Heat transfer unsteadiness relevance
While these recent developments have significantly advanced the geometric design and controllability of sweeping jets, most optimisation efforts remain rooted in time-averaged performance metrics. However, given the inherently unsteady nature of sweeping jets, a deeper understanding of the fluctuating heat transfer dynamics is essential for guiding more effective design strategies. Resolving the temporal behaviour of convective heat transfer, in particular, could uncover critical mechanisms responsible for enhanced mixing, localised cooling, and thermal uniformity—factors that directly influence the performance and reliability of optimised configurations. Although sweeping jets inherently exhibit large-scale unsteadiness due to their periodic lateral motion, the fluctuating heat transfer dynamics they induce have received comparatively limited attention. Most existing investigations have focused on time-averaged performance~\citep{kim2019measurement,gomes2024flow,dangelo2024piv}, which, although informative, offer only a partial understanding of the jet–surface interaction. By contrast, studies on other unsteady configurations, such as synthetic jets, have clearly demonstrated the importance of capturing instantaneous heat transfer variations to fully characterise and predict convective behaviour~\citep{greco2018effects}. These insights, derived from time-resolved heat transfer measurements, have proven essential not only for elucidating flow–surface coupling mechanisms, but also for optimising thermal performance and ensuring surface temperature uniformity~\citep{park2018flow,tu2023flow}.

Recent work by \citet{dangelo2025ir} has advanced the understanding of unsteady convective processes in sweeping jets through the use of phase-resolved infrared thermography. Their work employed a triple decomposition strategy to separate the thermal signal into time-averaged, coherent (phase-locked), and stochastic components. This method enabled the reconstruction of phase-averaged Nusselt number distributions, offering valuable insights into the periodic heat transfer behaviour associated with the sweeping cycle. However, as the data are conditioned on phase, this approach inherently smooths instantaneous fluctuations and therefore does not fully capture the local and instantaneous convective variations that occur during individual oscillation periods.

% Why PIRT?
In this context, \citet{PIRT_2025} recently demonstrated the importance of a non-phase-averaged approach, showing that, in rapidly oscillating and spatially complex flows, robust spatio-temporal filtering enables the extraction of quantitative fluctuating Nusselt number distributions that are directly linked to flow unsteadiness. Specifically, their study investigated the unsteady heat transfer dynamics of a sweeping jet in a configuration at $Re \approx 6600$ and $L/D = 2$. The results highlighted a significant contribution of unsteady effects to the Nusselt number distribution, thus motivating further investigations across a broader range of Reynolds numbers and nozzle-to-plate distances. Despite these advances, a comprehensive experimental assessment directly correlating unsteady wall heat transfer with flow dynamics in impinging sweeping jets remains lacking. Bridging this gap is crucial to fully understanding the coupled mechanisms that govern surface heat transfer and to supporting the development of more accurate models and optimised designs for advanced thermal management applications.

% What we propose:
Motivated by these developments, the present study aims to advance the understanding of sweeping jet impingement by providing a comprehensive experimental characterisation of both flow and surface heat transfer fluctuations. Building upon previous heat transfer~\citep{dangelo2025ir} and flow field~\citep{dangelo2024piv} studies, and adopting the advanced data processing methodology proposed by~\citet{PIRT_2025}, the current work integrates non-synchronous planar PIV and time-resolved infrared thermography to directly relate the time-averaged and fluctuating convective heat transfer at the impingement surface with the jet dynamics. The oscillatory nature of sweeping jets induces spatially localised thermal unsteadiness, which can significantly impact thermal stresses, temperature uniformity, and overall cooling efficiency in thermal management systems. This study aims to bridge this gap by providing a comprehensive experimental analysis of both time-averaged and fluctuating Nusselt number distributions in sweeping jet impingement, focusing on how Reynolds number and nozzle-to-plate distance influence these key heat transfer characteristics. Hence, the research questions are twofold: 
\begin{itemize}
    \item \textit{What is the correlation between the Nusselt number on the wall and the velocity field?}
    \item \textit{Can we identify a reduced-order model for heat transfer that highlights the relevance of key operating parameters such as Reynolds number and nozzle-to-plate distance?}
\end{itemize}
To answer these questions, the proposed experiments cover nozzle-to-plate distances from $L/D = 1$ to $5$ and Reynolds numbers up to approximately $\Ren = 12000$, enabling a systematic assessment of the coupled unsteady flow and thermal response under realistic operating conditions. Unlike previous studies, which focused separately on either the flow field or the heat transfer response, often at a single operating point, the present work explores their coupled behaviour across a wide parametric space of $\Ren$ and $L/D$.  
Finally, we propose a data-driven surrogate model based on manifold embedding, offering a predictive tool to estimate convective heat transfer response across varying flow conditions, providing valuable insights for the optimisation of cooling systems in complex thermal environments. To the best of the authors' knowledge, this is the first study to embed such combined thermal and flow data into a low-dimensional parametric space, yielding a compact, low-dimensional representation of sweeping jet behaviour, suitable for integration into early-stage design and optimisation frameworks.

The remainder of this paper is structured as follows: \autoref{s:Methodology} details the experimental setup and methodology. The results are presented and discussed in \autoref{s:results}, followed by the main conclusions in \autoref{s:Conclusions}.

%\afterpage{\clearpage} %Nomenclature forced
%%%%%%%%%%%%%%%%%%%%%%%%%%%%%%%%%%%%%%%%%%%%%%%%%%%%%%%%%%%%%%%%%%%%%
%%%%%%%%%%%%%%%%%%%%%%%%%%% METHODOLOGY %%%%%%%%%%%%%%%%%%%%%%%%%%%%%
%%%%%%%%%%%%%%%%%%%%%%%%%%%%%%%%%%%%%%%%%%%%%%%%%%%%%%%%%%%%%%%%%%%%%
\section{Experimental setup and Methodology \label{s:Methodology}}
%
%-Setup Image
\begin{figure*}[t]
    \centering
    \includegraphics[width=0.9\linewidth]{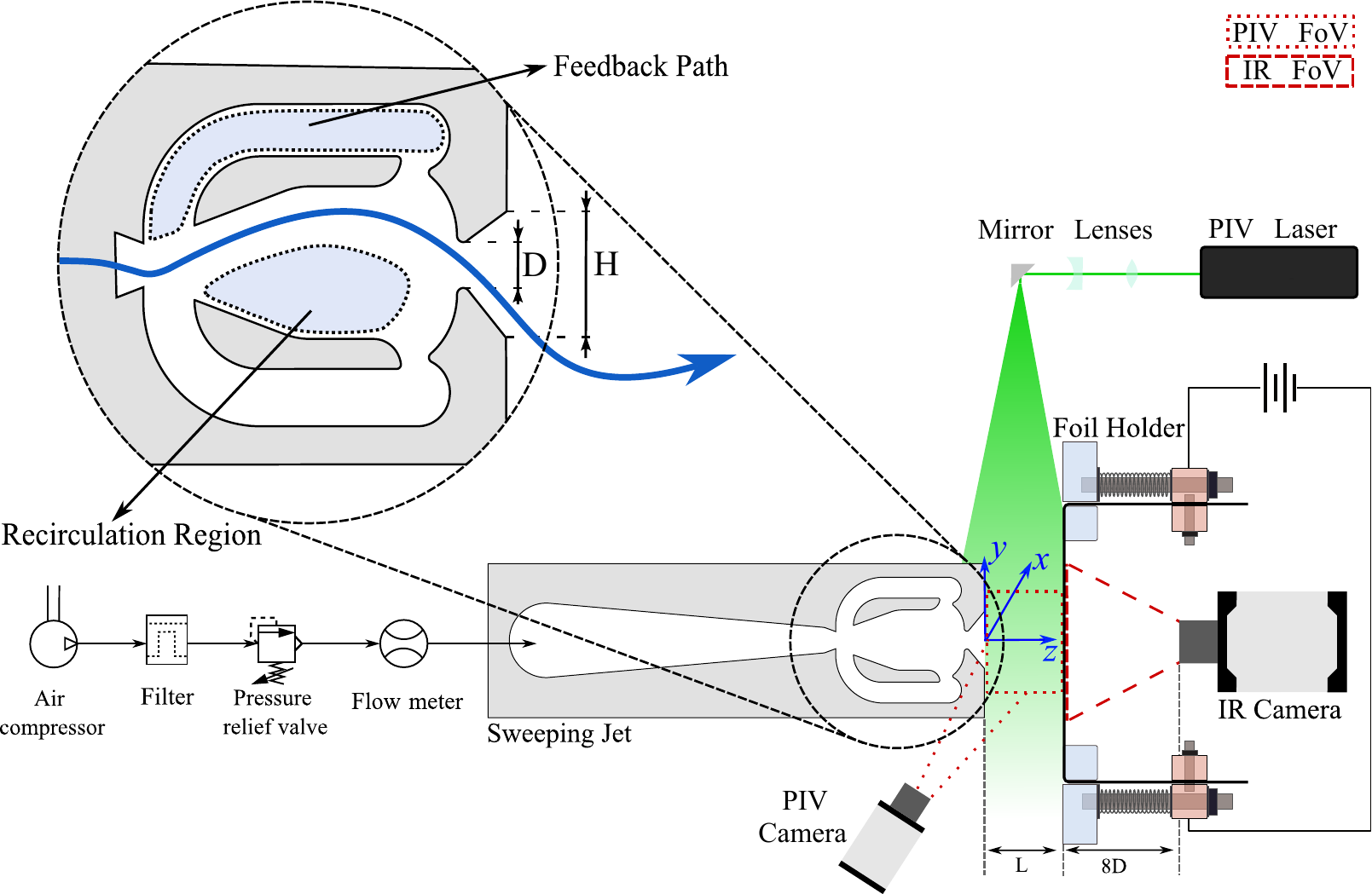}
    \caption{Schematic of the experimental setup combining the sweeping jet impingement configuration with infrared thermography and planar PIV measurements. The impinging jet is generated by a fluidic oscillator with a throat width of $D = 6.25$~mm, installed within a channel of height $H = 20$~mm, a geometry taken from \citep{kim2019measurement}. The nozzle-to-plate distance $L$ is varied to investigate its effect on flow and heat transfer. The target plate consists of a heated thin foil sensor monitored by a time-resolved IR camera, while a PIV system (laser, optics, and camera) captures the velocity field in the spanwise--wall-normal ($y$--$z$) plane intersecting the jet midline. The PIV and IR fields of view (FoV) are indicated for reference. Non-synchronous acquisition avoids thermal artifacts from laser exposure during PIV operation. Figure adapted from the schematic representation in \citet{PIRT_2025}.}
    \label{fig:setup}
\end{figure*}
Impinging jet experiments were conducted using a fluidic oscillator with the same internal geometry as described by \citet{park2018flow}. The device features a square throat area with a width ($D$) of 6.25 mm and an exit angle of $100^{\circ}$. A schematic of the geometry is shown in \autoref{fig:setup}, highlighting the nozzle throat ($D$), the channel height ($H$), and the nozzle-to-plate distance ($L$). 

The oscillator was manufactured from transparent methacrylate, with a wall thickness of 20 mm to ensure structural integrity and optical access.  Compressed air was supplied through a regulated line equipped with an Alicat Scientific\textsuperscript{TM} M-500SLPM flow controller. This device precisely regulates the mass flow rate and simultaneously monitors key thermodynamic properties, including absolute pressure, temperature, and volumetric flow rate at the inlet.

\begin{figure}
    \centering
    \includegraphics[width=0.99\linewidth]{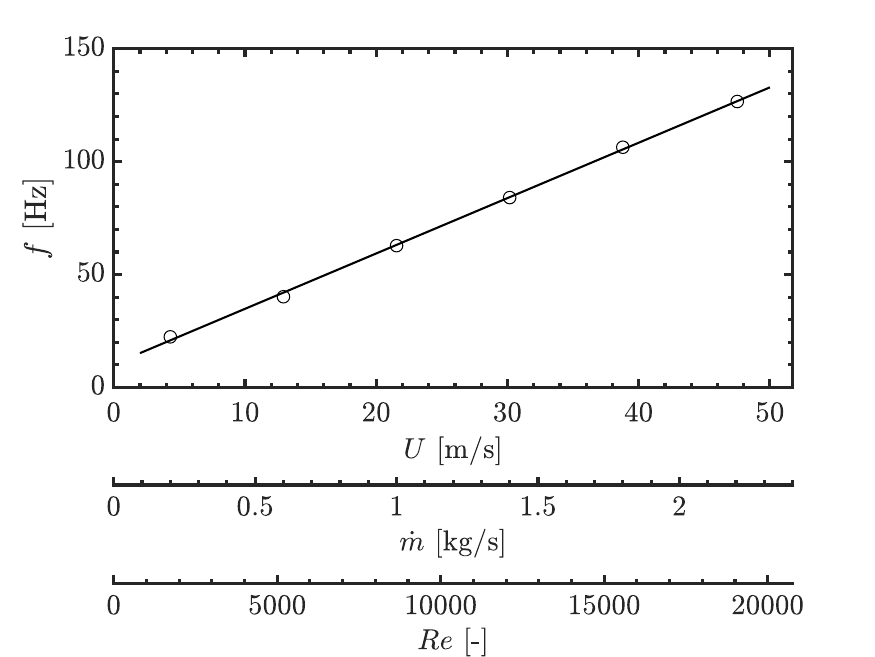}
  \caption{Sweeping frequency $f_s$ of the fluidic oscillator as a function of the average jet velocity $U$ (bottom axis), mass flow rate $\dot{m}$ (middle axis), and Reynolds number $\Ren$ (top axis). The linear correlation confirms that the Strouhal number remains constant across the tested range, with $St \approx 0.0625$. Data support the scaling laws typically reported for oscillating jets generated by fluidic oscillators.}
    \label{fig:F1_Freq}
\end{figure}

A characterisation of the sweeping frequency ($f_s$) of the fluidic oscillator was conducted prior to the experimental campaign and is reported in \autoref{fig:F1_Freq}.  The frequency was estimated by recording the acoustic signal of the jet using a MEMS microphone (Knowles SPH0645LM4H-B, 44.1~kHz sampling rate) placed at a fixed location downstream of the nozzle in a free jet configuration (i.e., without the impingement plate). A 30-second recording was analysed via Fast Fourier Transform, revealing the dominant sweeping frequency at each flow rate. The data confirm that $f_s$ increases linearly with the mass flow rate ($\dot{m}$), namely the bulk velocity at the nozzle exit ($U$) or the Reynolds number ($\Ren = {UD}/{\nu}$).  This behaviour is consistent with previous studies on similar oscillator geometries, where the sweeping mechanism is known to scale with the momentum flux through the oscillator~\cite{woszidlo2019fundamental,Bobusch2013internalSJ}.  The resulting Strouhal number ($St = {f_s D}/{U}$) remains approximately constant, with $St \approx 0.0625$, across the range of operating conditions.  This constant $St$ indicates that the oscillation dynamics for an incompressible regime are governed predominantly by the internal geometry of the fluidic oscillator, rather than by external flow conditions \citep{Schmidt2017separation}.

Experiments were performed by varying both the mass flow rate and the nozzle-to-plate distance. Specifically, $\dot{m}$ was varied from 0.2 to 1.4 g/s, corresponding to Reynolds numbers in the range $\Ren = 1750$–$12270$. This range was chosen to encompass both transitional and turbulent flow regimes, thereby enabling the investigation of how the organised side-to-side jet sweeping interacts with the intrinsic turbulent fluctuations that grow with Re. The lower bound ensures a sufficiently low $\mathrm{Gr}/\Ren^2$ ratio to minimise buoyancy effects and maintain forced convection dominance, while the upper bound is constrained by the temporal resolution limit of the heated-thin-foil sensor and the IR camera (see Eqs.~\ref{eq:Fo} and~\ref{eq:SNR}, respectively). The normalised nozzle-to-plate spacing ($L/D$) was modified between 1 and 5, a range consistent with previous studies on impinging jets~\cite{dangelo2024piv,dangelo2025ir}, and selected to capture the influence of geometric confinement on the development of the sweeping motion and its thermal footprint. Smaller $L/D$ values promote early jet impingement and potential stagnation effects, whereas larger spacings allow for greater lateral jet deflection and full development of the sweeping structure before impingement. 

Heat transfer and flow field measurements were conducted using the experimental setup illustrated in the schematic in \autoref{fig:setup}. The impinged surface consists of a custom-built heated thin-foil sensor, allowing for spatially and temporally resolved measurements of wall heat flux via infrared thermography.
Two complementary diagnostic systems were deployed: a time-resolved IR camera (silver housing) to capture the surface temperature distribution, and a planar PIV system (blue camera and black laser head) to resolve the velocity field in the spanwise–wall-normal plane. The laser sheet intersects the centreline of the sweeping jet and illuminates tracer particles seeded into the flow. The two measurement techniques were operated in a non-synchronous manner, allowing independent acquisition of velocity and thermal data while avoiding spurious heating of the thin foil due to laser-induced absorption during PIV. While temporal correlation between flow and thermal data is not available, the non-simultaneous acquisition enables a consistent comparison of statistical quantities, such as time-averaged fields and root-mean-square fluctuations, under the same flow conditions.
Details on the calibration, uncertainty, and data processing of each technique are provided in the following subsections.

%---------------------------------------------------------------------------
%%%%%%%%%%%%%%%%%%%%%%%%%%%%%%%%%%%% IR %%%%%%%%%%%%%%%%%%%%%%%%%%%%%%%%%%%%
\subsection{Time-resolved heat transfer measurements \label{ss:IR}}
The wall heat transfer distribution is measured using a heated thin-foil heat-flux sensor coupled with high-speed infrared thermography, following the approach developed in previous studies~\citep{mallor2019modal, Castellanos2022slotjet, PIRT_2025}. The sensor consists of a $10,\mu$m-thick stainless steel foil ($\delta_f$), mounted flush with the target wall using a custom-designed 3D-printed frame made of polylactide (PLA). The foil spans an area of $100 \times 250,\mathrm{mm}^2$, and is supported by two thermally insulated aluminum rods, preventing parasitic heat conduction from the frame. The foil is tensioned by spring-loaded copper clamps, which also serve as electrical contacts. Indium wire ($1,\mathrm{mm}$ thick) is embedded in the clamp interface to ensure uniform current distribution and minimise contact resistance.

Surface temperatures were recorded using a TELOPS TS-IR MW-IN camera equipped with a $640 \times 512$ pixel Mercury–Cadmium–Telluride detector and a noise-equivalent temperature difference (NETD) below $20,\mathrm{mK}$. The acquisition rate was set to $310,\mathrm{Hz}$, and the spatial resolution was approximately $2.0,\mathrm{pixels/mm}$. The selected spatial and temporal resolutions ensured that both the spatial gradients $\nabla^{2}T$ and temporal derivatives $\partial T / \partial t$ were captured with sufficient fidelity, with any related uncertainties fully accounted for in the uncertainty analysis. To ensure measurement accuracy, a two-point radiometric calibration was performed \textit{ex situ}, replicating the optical path used during the experiments. The foil was coated on both sides with a high-emissivity paint ($\varepsilon \approx 0.95$), resulting in a total coating thickness of $\delta_p = 42,\mu\mathrm{m}$. The paint thickness was measured using a micrometre, and its thermal properties were estimated based on average values reported by \citet{PaintConvection_Raghu_2006}, following the recommendations of \citet{PaintConvection_Stafford2009}.

Joule heating is induced by supplying a stabilised electric current across the copper clamps using a programmable power source (BK PRECISION 1901B). The input heat flux is given by $q''{j} = V{DC} I / A_f$, where $V_{DC} \approx 1.9,\mathrm{V}$ and $I \approx 7.5,\mathrm{A}$ are the voltage and current, respectively. The power is constant and spatially uniform due to the foil’s homogeneous resistance and uniform contact.

The local convective heat transfer coefficient $h$ is calculated from an unsteady energy balance on the sensor:
%----------------------------------------------------------------------------
\begin{equation}
	 h = \frac{ q''_{j} - 2 q''_{r} - q''_{k} - q''_{n} -(\rho_{f}c_{f}\delta_f+\rho_p c_p \delta_p)\frac{dT_w}{dt} }{ T_\mathrm{w} - T_\mathrm{aw} } ,
	\label{eq:heatedthinfoil}
\end{equation}
%----------------------------------------------------------------------------
where $q''_r$ denotes the radiative heat flux from both surfaces of the foil, $q''_k$ corresponds to the in-plane (tangential) conduction losses across the foil surface, and $q''_n$ accounts for natural convection losses from the sensor’s back side. The temperatures $T_\mathrm{w}$ and $T_\mathrm{aw}$ represent the actual wall temperature and the adiabatic wall temperature, respectively, obtained from measurements with and without Joule heating, as later explained. Radiative losses are estimated by assuming the surroundings behave as a black body at ambient temperature, leading to $q''_r = \sigma \varepsilon (T_\mathrm{w}^4 - T_\infty^4)$, where $\sigma$ is the Stefan–Boltzmann constant and $\varepsilon$ is the emissivity of the foil surface with the high-emissivity coating. Tangential conduction losses are calculated as $q''_{k} = (k_f \delta_f + k_p \delta_p) \nabla^{2} T_\mathrm{w}$, based on the Laplacian of the surface temperature and the combined thermal resistance of the foil and paint layers. Natural convection on the flow-facing side of the foil is neglected, as $\mathrm{Gr/Re}^2<< 1$, where $\mathrm{Gr} = {a_g\beta_v(T_\mathrm{w}-T_\mathrm{aw}) D^3}/{\nu^2}$ is the Grashof number based on the nozzle exit width. On the back side of the sensor, $q''_n$ is estimated using standard correlations for natural convection over vertical plates\citep{bergman2011fundamentals}, and is found to remain below $1\%$ of the applied Joule heating across all tested conditions.

In the context of unsteady convective heat transfer, particularly under oscillatory conditions such as those induced by sweeping jets, capturing the fluctuating heat flux at the wall is essential. To resolve these unsteady thermal effects, we employ a heated thin-foil sensor coupled with time-resolved infrared thermography. The measurement process consists of two sequential acquisitions: one under Joule heating (\textit{hot} image), providing the surface temperature distribution $T_\mathrm{w}$, and one without electrical input (\textit{cold} image), yielding the adiabatic wall temperature $T_\mathrm{aw}$. The hot and cold sequences comprise 10000 and 1000 snapshots, respectively, to ensure adequate statistical convergence. Although the infrared camera views the rear face of the foil, the temperature across its thickness can be considered uniform. This assumption is validated by the low Biot number, defined as:
\begin{equation}\label{eq:Bi} 
    \mathrm{Bi} = \frac{h s_{eq}}{k_{eq}} = h \left(\frac{s_f}{k_f} + \frac{s_p}{k_p} \right), 
\end{equation}
where $s_f$ and $s_p$ are the thicknesses of the foil and the high-emissivity paint layer, respectively, and $k_f$, $k_p$ their corresponding thermal conductivities. Across the entire test cases, the Biot number remains in the range $10^{-5} < \mathrm{Bi} < 10^{-3}$, ensuring one-dimensional conduction and thermal uniformity within the foil.

To resolve unsteady effects, the sensor must also exhibit fast thermal response relative to the characteristic timescale of the heat flux fluctuations. This is quantified via the Fourier number:
\begin{equation}\label{eq:Fo} 
    \mathrm{Fo} = \frac{\alpha_{eq}}{ f s_{eq}^2} = \frac{1}{f \left(\frac{s_f}{k_f}+\frac{s_p}{k_p}\right)\left[s_f (\rho c)_f + s_p (\rho c)_p\right]}, 
\end{equation}
which incorporates the thermal resistance and inertia of both the foil and the surface coating. For current experimental conditions, the Fourier number varies with the oscillation frequency, which scales with the Reynolds number, yielding $1.9 < \mathrm{Fo} < 7.2$ in the tested range. The limiting case of $\mathrm{Fo} \approx 1.9$ may result in partial attenuation of the fluctuating Nusselt number signal due to insufficient thermal penetration depth. This subtle limitation must be taken into account when interpreting the slight deviation from the monotonic increase in $\Nu'$ with $\Ren$, as discussed in Section~\ref{ss:heattransfer}.

In addition to satisfying the Biot and Fourier criteria, the system's thermal response must also exceed the infrared camera’s sensitivity threshold. This is evaluated by comparing the magnitude of the surface heat flux fluctuations with the camera’s noise-equivalent temperature difference (NETD). A minimum signal-to-noise ratio is required:
\begin{equation} \label{eq:SNR}
\frac{ q''_{j} - 2 q''_{r} - q''_{k} - q''_{n} }{(\rho_{f}c_{f}\delta_f + \rho_p c_p \delta_p) \cdot \mathrm{NETD} } \gg 1, 
\end{equation}
where the numerator accounts for the net heat input after subtracting radiative losses, conduction, and noise terms. This ensures that thermal fluctuations induce detectable surface temperature changes above the NETD threshold. For further detail on the derivation of the maximum measurable frequency and the signal-to-noise constraints for IR-based heat transfer sensors, the reader is referred to~\citep{nakamura2009frequency,FCutStafford2012}.

The adequacy of Biot and Fourier numbers as well as the NETD threshold assessment was evaluated pixelwise for each case, ensuring that the condition is satisfied for all the images and proving a good uniformity of the IR images. Additionally, before testing, the spatial uniformity of the heating foil was evaluated under low-power steady-state conditions without jet impingement. The temperature field was found to be uniform across the surface, with only minor deviations due to residual natural convection. These deviations disappear during forced convection testing, even at the lowest Reynolds number investigated, as highlighted by the low ratio between the Grashof number and the squared Reynolds number. Tangential conduction effects within the foil and paint layers are accounted for through the term $q''_k$ included in Eq.~\eqref{eq:heatedthinfoil}. In practice, the values of $q''_k$ were found to be negligible and primarily composed of noise, consistent with the findings reported in~\cite{PIRT_2025}. For this reason, $q''_k$ is not presented in the results but remains included in the model to ensure robustness.

Once the infrared sequences are acquired, the temperature fields are not directly processed through the energy balance equation. Instead, a dedicated spatio-temporal filtering pipeline is applied to ensure a physically consistent extraction of unsteady heat transfer signals. The filtering strategy is based on the methodology introduced by \citet{PIRT_2025}, which demonstrated the importance of suppressing acquisition noise, optical artifacts, and digitisation effects for accurate interpretation of high-speed thermal data.

The first step is a modal decomposition via Proper Orthogonal Decomposition (POD), which serves as a denoising filter by removing the least energetic modes of the thermal signal. This technique preserves the dominant coherent structures while minimising the contribution of sensor noise and low-energy background fluctuations. The filtering is performed by truncating the POD expansion, and the number of modes retained is determined via the Optimal Hard Threshold criterion, which minimises the worst-case error between the filtered and true signals \citep{gavish2014optimal}. A second layer of filtering is applied through a two-dimensional Gaussian convolution, which mitigates residual spatial noise and corrects for sensor-induced artefacts such as non-uniformity correction (NUC) errors. The use of a localised Gaussian kernel maintains the integrity of thermal gradients while suppressing pixel-level outliers. Finally, a Savitzky–Golay filter is used in the temporal direction, which provides a smooth reconstruction of thermal evolution and allows for accurate calculation of temporal derivatives—crucial for the evaluation of the unsteady term in the energy balance. This multi-tiered filtering approach is essential for reliably isolating the convective contribution to the surface heat transfer in oscillating flows, and ensures a smoother and cleaner Nusselt number spatial map, effectively reducing noise that could distort temporal derivatives and, consequently, the fluctuating component of heat transfer. Although maximum values may be attenuated due to slightly lower spatial and temporal resolution compared to previous studies~\citep{PIRT_2025}, this approach was prioritised for its ability to maintain a clearer representation of the thermal footprint, which is crucial for the reliable interpretation and analysis of the unsteady dynamics. This represents the first reporting of these fields, and clarity in spatial distribution is essential for accurate data-driven modelling and understanding of the convective heat transfer processes.

The resulting heat transfer coefficient $h$ is then expressed in dimensionless form using the Nusselt number, defined as $\Nu = h D / k_{\mathrm{air}}$, where $k_{\mathrm{air}}$ is the thermal conductivity of air evaluated at the film temperature $(T_w + T_{aw})/2$. This formulation enables a consistent comparison across different flow conditions and geometric configurations and facilitates integration with the velocity-based analysis.

\begin{table}
\centering
\caption{Uncertainty contributions for Nusselt number calculation.} \label{tab:uncertainties}
    \begin{tabular*}{\tblwidth}{@{}LLL@{}}
        \toprule
        Parameter               & Uncertainty   & Typical Value \\ \midrule
        $T_\mathrm{w}$          & 0.1 K         & 310 [K]  \\
        $T_\mathrm{aw}$         & 0.1 K         & 300 [K] \\
        $T_\infty$              & 0.1 K         & 305 [K] \\
        $V_{DC}$                & 0.2\%         & 7.5 [V] \\
        $I$                     & 0.2\%         & 1.9 [A] \\
        $\varepsilon$           & 2\%           & 0.95 \\
        $A_f$                   & 0.1\%         & 250 [cm$^2$] \\
        $\delta_f$              & 5\%           & 10.0 [$\mu$ m] \\
        $\delta_p$              & 5\%           & 42.0 [$\mu$ m] \\
        $k_{{\mathrm{air}}}$    & 1\%           & $1.47\cdot 10^{-4}$ [W/k] \\
        $k_{{\mathrm{f}}}$      & 1\%           & $5.79\cdot 10^{-5}$ [W/k] \\
        $k_{{\mathrm{p}}}$    & 6.12\%        & 1.38 [W/(m K)] \\
        $c_{{\mathrm{f}}}$      & 2\%           & 460 [J/(kg K)] \\
        $c_{{\mathrm{p}}}$      & 10\%           & 3061.5 [J/(kg K)] \\
        $\rho_{{\mathrm{f}}}$   & 1\%           & 7900 [$kg/m^3$)] \\
        $\rho_{{\mathrm{p}}}$   & 3.25\%           & 1261.2 [$kg/m^3$)] \\
        $\frac{\partial T_w}{\partial t}$       & 10\%          &  1.2 [K/s] \\
        $\frac{\partial^2 T_w}{\partial x^2}$       & 10\%          & 1259.6 [$k^2/m^2$] \\
        $\frac{\partial^2 T_w}{\partial y^2}$       & 10\%          & 644.2 [$k^2/m^2$] \\
        $D$                     & 1\%           & 6.25 [cm] \\
        \bottomrule
    \end{tabular*}
\end{table}

The overall uncertainty associated with the Nusselt number is evaluated through a Monte Carlo simulation procedure, assuming statistically independent sources of error and incorporating the uncertainties listed in Table~\ref{tab:uncertainties}, in accordance with established IR measurement protocols~\citep{minkina2009infrared,Castellanos2022slotjet,PIRT_2025}. The resulting uncertainty on $\Nu$ remains below $\pm 6\%$ throughout the entire test matrix, confirming the robustness of the acquisition and post-processing methodology. In addition, to ensure the repeatability of the heat transfer measurements, two independent experimental runs were conducted for each configuration, with a third run performed if discrepancies above uncertainty were detected. The Nusselt number distributions from these repeated runs showed consistent results within the uncertainty bounds, validating the reliability of the measurements.

%----------------------------------------------------------------------------
%%%%%%%%%%%%%%%%%%%%%%%%%%%%%%%%%%% PIV  %%%%%%%%%%%%%%%%%%%%%%%%%%%%%%%%%%%
\subsection{Velocity measurements \label{ss:PIV}}
The heat transfer analysis is complemented by flow field measurements obtained using two-component Particle Image Velocimetry (PIV), which provides detailed velocity information in the centerplane ($y/L = 0$), as illustrated in Figure~\ref{fig:setup}. The imaging system consists of an iLA5150 sCMOS camera equipped with a $50,\mathrm{mm}$ focal-length lens and a focal ratio of $f/\# = 11$, yielding a spatial resolution of approximately $40 \mathrm{pixels/mm}$. The field of view is cropped to capture the region surrounding the impinging jet and the wall interaction zone.
The flow is seeded with Di-Ethyl-Hexyl-Sebacate tracer particles of approximately $1\mu\mathrm{m}$ diameter. Illumination is provided by a dual-cavity Nd:YAG Quantel Evergreen laser, delivering $200 \mathrm{mJ}$ per pulse at $15 \mathrm{Hz}$. The laser beam is shaped into a light sheet using a combination of cylindrical and spherical lenses. A total of 1100 image pairs are acquired at $15 \mathrm{Hz}$ to ensure statistical convergence of the flow quantities.

Preprocessing of the raw images includes background subtraction using the POD-based technique introduced by \citet{Mendez2017pod-piv}, which enhances contrast and reduces static artefacts. Image cross-correlation is performed using the in-house software developed by the Experimental Thermo-Fluid-Dynamics Group at the University of Naples Federico II. The algorithm employs a multi-pass, window-deformation approach~\citep{soria1996piv, scarano2001iterativeimgdef}, with advanced weighting functions and interpolation schemes to optimise spatial resolution and vector precision~\citep{Astarita2006PIV, Astarita2007PIV}. The final interrogation window size is set to $48 \times 48 \mathrm{pixels}^2$ with 75\% overlap, resulting in a vector spacing of approximately $1.4 \mathrm{vectors/mm}$. The estimated uncertainty in the displacement field is $\sim 0.1$ pixels~\citep{raffel2018piv}. %A small time separation ($20\mu s$) between the laser impulses was set to improve the signal-to-noise ratio in the correlation process, accounting for the low seeding density (the jet was not seeded) and large velocity gradients. This resulted in typical displacements of $4$ pixels, leading to an uncertainty of about 2.5\% on the instantaneous velocity fields.

%%%%%%%%%%%%%%%%%%%%%%%%%%%%%%%%%%%%%%%%%%%%%%%%%%%%%%%%%%%%%%%
%%%%%%%%%%%%%%%%%%%%%%%%%%% RESTULS %%%%%%%%%%%%%%%%%%%%%%%%%%%
%%%%%%%%%%%%%%%%%%%%%%%%%%%%%%%%%%%%%%%%%%%%%%%%%%%%%%%%%%%%%%%
\section{Heat Transfer and Flow Field Analysis \label{s:results}}

This section presents a detailed analysis of the flow and heat transfer characteristics of the impinging sweeping jet under varying Reynolds numbers and nozzle-to-plate distances. Time-averaged and fluctuating components of the thermal response are first examined through infrared thermography. These are then interpreted in light of velocity field measurements obtained via PIV, offering a physical explanation for the spatial organisation of convective transport. Finally, the relationship between flow unsteadiness and thermal fluctuations is quantified, consolidating the coupling between jet dynamics and wall heat transfer.

\subsection{Surface Heat Transfer distribution \label{ss:heattransfer}}
\begin{figure*}
    \centering
    \includegraphics[width=0.99\linewidth]{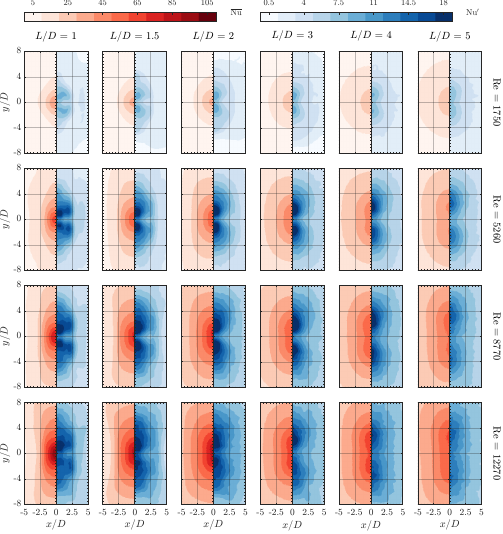}
    \caption{Spatial distribution of time-averaged Nusselt number ($\overline{\Nu}$, left half of each panel) and its fluctuating component ($\Nu'$, right half) for different Reynolds numbers and nozzle-to-plate distances ($L/D$). The left-to-right direction corresponds to increasing $L/D$, while top-to-bottom represents increasing $\Ren$. The maps reveal the effects of jet oscillation, impingement footprint spreading, and fluctuating heat transfer, particularly at high Reynolds numbers and moderate $L/D$.}
    \label{fig:F1_Numaps}
\end{figure*}

% Maps:
Figure~\ref{fig:F1_Numaps} presents the spatial distributions of both the time-averaged Nusselt number, $\overline{\Nu}$ (left side of each subpanel), and its fluctuating component, $\Nu'$ (right side), over the impingement surface for various combinations of Reynolds number ($\Ren$) and nozzle-to-plate distance ($L/D$). Each column corresponds to a different $L/D$ ratio, while each row represents a fixed Reynolds number. The results provide a comprehensive view of how both the mean and unsteady heat transfer evolve with flow conditions and geometric parameters. 

The mean Nusselt number consistently increases with $\Ren$, indicating enhanced convective transport associated with larger jet momentum and higher oscillation frequencies. At higher $\Ren$, the jet impinges more forcefully on the surface, increasing turbulence levels and promoting more effective fluid–surface energy exchange. The spatial distribution of $\overline{\Nu}$ shows a peak in the central region near the stagnation point, gradually decaying toward the periphery. However, as $L/D$ increases, this footprint widens, particularly in the $y$-direction, which corresponds to the plane of jet oscillation, in agreement with previous results reported in \citep{zhou2019heat}. This widening is attributed to the extended sweep length of the jet at larger impingement distances. When the wall is closer to the nozzle (low $L/D$), the oscillation is geometrically constrained, resulting in a narrow footprint with a higher localised maximum. At larger $L/D$, although the footprint expands, the peak $\overline{\Nu}$ diminishes and the overall distribution becomes more uniform due to increased jet dispersion and reduced impact strength.

The fluctuating Nusselt field ($\Nu'$), shown on the right side of each subpanel, offers additional insights into the unsteady heat transfer dynamics. These distributions reveal regions of significant thermal fluctuation induced by the jet’s sweeping motion, so that $\Nu'$ exhibits a distinct double-lobe structure aligned along the $y$-axis. These off-center peaks emerge due to the periodic impingement of the jet in alternating lateral positions, where the local wall shear and heat transfer rates become intermittently enhanced. This feature, which cannot be captured from $\overline{\Nu}$ alone, is a direct result of the oscillating nature of the jet and agrees with previous Lagrangian-based interpretations of sweeping jet dynamics~\citep{wen2018lagrangian}.
The emergence and spacing of these $\Nu'$ peaks become more pronounced as $\Ren$ increases, reflecting a broader sweep angle and higher amplitude oscillations. A similar trend is observed with increasing $L/D$, though at very large spacings (e.g., $L/D = 5$), the fluctuating footprint becomes more diffuse and less intense, suggesting a decay in oscillation strength due to dissipation in the free jet region.

It is important to highlight that the accurate estimation of $\Nu'$ critically depends on appropriate spatio-temporal filtering of the infrared data, as discussed in~\citet{PIRT_2025}. In the absence of such filtering, unsteady thermal contributions may be artificially attenuated, obscuring the characteristic double-peak structure or underestimating fluctuation intensity. In this regard, the present study also aims to underscore the significance of the unsteady term in the thin-foil energy balance equation, which is often neglected in simplified analyses. Although the fluctuating component may represent only a fraction of the mean Nusselt number, our results show that $\Nu'$ can locally reach values up to 10-18\% of $\overline{\Nu}$, especially near the lateral sweep edges, highlighting its non-negligible contribution to the overall convective transport process.

Additionally, although all configurations exhibit a Fourier number $\mathrm{Fo}$ above unity, indicating an acceptable thermal response by the sensor, the highest Reynolds number cases approach the lower resolution limit ($\mathrm{Fo} \approx 1.9$). Such limitation suggests that the performance of the heat-flux sensor may no longer be sufficient to fully resolve the rapid dynamics of the sweeping jet, potentially leading to a partial attenuation of the $\Nu'$ signal. This effect is consistent with the observed deviation from the otherwise monotonic increase in $\Nu'$ magnitude with $\Ren$.

The spatial distributions presented in \autoref{fig:F1_Numaps} further underscore the importance of analysing both the time-averaged and fluctuating components of the surface heat transfer field to fully characterise the dynamics of impinging sweeping jets. While $\overline{\Nu}$ maps provide insight into the average cooling effectiveness, the fluctuating component $\Nu'$ captures the signature of unsteady jet–surface interactions that arise from the oscillatory behaviour of the flow. These unsteady interactions are critical to understanding the spatio-temporal evolution of convective heat transfer and are particularly relevant for applications involving flow control, surface wear, or thermal fatigue, where transient phenomena dominate the surface response.

% Integral Profiles:
To quantitatively assess the global performance of the sweeping jet over the impingement surface, we define a figure of merit based on the spatial average of the Nusselt number over a growing region of interest centered on the stagnation zone. Given that the heat transfer enhancement tends to be spatially localised and highly dependent on the jet parameters, we systematically vary the size of this region to evaluate how efficiently the jet delivers heat transfer enhancement per unit area. The region of integration $\mathcal{D}_\beta$ is defined as a rectangular area that scales with a dimensionless parameter $\beta = x/D$, where $D$ is the hydraulic diameter of the jet and $x$ denotes the half-width of the domain along the wall-parallel direction. The vertical extent of the rectangle is given by $y = \chi \cdot x$, with $\chi = H/D = 6.5/20$ being the aspect ratio of the nozzle. This ensures that the integration domain retains the same geometric proportions as the exit cross-section of the jet. Over each region $\mathcal{D}_\beta= \{(x,y) \in \mathbb{R}^2:  -\beta \leq x/D \leq \beta; y=\chi \cdot x; \beta \in [0,5]\}$, we compute the integral average of a generic scalar field $\Theta$ (such as $\overline{\Nu}$ or $\Nu'$): 
\begin{equation} \label{eq:Num} \langle \Theta \rangle = \frac{1}{m(\mathcal{D}_\beta)}\iint_{\mathcal{D}_\beta} \Theta(x,y) \,dx\,dy, \end{equation}
and the associated spatial standard deviation, which quantifies the degree of spatial uniformity of the heat transfer enhancement within that region:
\begin{equation} \label{eq:Numstd} \sigma_\Theta =  \left( \frac{1}{m(\mathcal{D}_\beta)} \iint_{\mathcal{D}_\beta} \left(\Theta(x,y)-\langle \Theta \rangle\right)^2 \,dx\,dy \right)^{1/2} \end{equation}

\begin{figure*}
    \centering
    \includegraphics[width=0.99\linewidth]{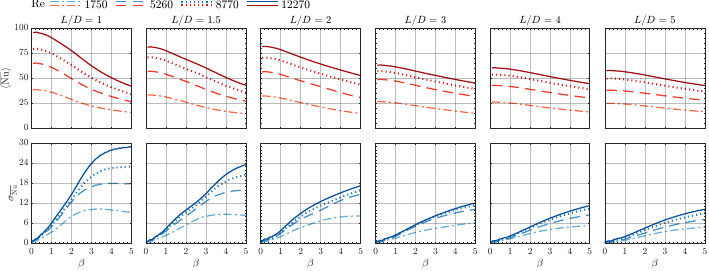}
    \caption{Spatial average of the time-averaged Nusselt number, $\langle \overline{\Nu} \rangle$, and its standard deviation, $\sigma_{\overline{\Nu}}$, as a function of the normalised impingement area size $\beta$. Each subplot corresponds to a different $L/D$ ratio, and curve colour intensity increases with Reynolds number.}
    \label{fig:F2_Num}
\end{figure*}

\begin{figure*}
    \centering
    \includegraphics[width=0.99\linewidth]{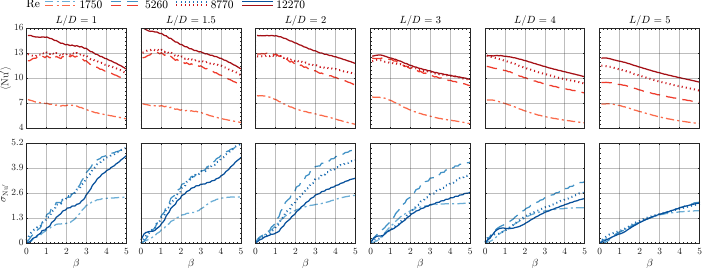}
    \caption{Spatial average of the fluctuating component of the Nusselt number, $\langle \Nu' \rangle$, and its standard deviation, $\sigma_{\Nu'}$, as a function of the normalised impingement area size $\beta$. Each subplot corresponds to a different $L/D$ ratio, with darker curves representing higher Reynolds numbers.}
    \label{fig:F3_Nurms}
\end{figure*}

This analysis is repeated for increasing values of $\beta$, effectively exploring how the performance and concentration of the heat transfer evolve with the size of the impingement area. By tracking both the average and the spread (standard deviation) of $\overline{\Nu}$ and $\Nu'$, we gain insight into how localised or diffused the thermal footprint becomes under different operating conditions. \autoref{fig:F2_Num} and~\autoref{fig:F3_Nurms} summarise the evolution of the spatially averaged Nusselt number and its standard deviation for both the mean ($\overline{\Nu}$) and fluctuating ($\Nu'$) components as a function of the integration area size, defined by the normalised extent $\beta$. Each subplot corresponds to a different nozzle-to-plate spacing $L/D$, while the curve color intensity represents the Reynolds number: darker curves denote higher $\Ren$.

\autoref{fig:F2_Num} shows that $\langle \overline{\Nu} \rangle$ decreases monotonically with $\beta$ for all $L/D$ values. This behavior is expected, as the regions closest to the stagnation point exhibit the highest heat transfer, while extending the averaging area progressively includes peripheral zones with lower $\overline{\Nu}$ values. As the Reynolds number increases, $\langle \overline{\Nu} \rangle$ also increases consistently, indicating that enhanced jet momentum improves convective performance even over larger surface areas. The rate at which $\langle \overline{\Nu} \rangle$ decreases with $\beta$ varies with $L/D$. For small nozzle-to-plate distances, the decay is sharper, reflecting a more concentrated thermal footprint. Conversely, at larger $L/D$, the decline is milder, consistent with a broader and more uniformly distributed heat transfer pattern.

The associated standard deviation, $\sigma_{\overline{\Nu}}$, increases with $\beta$ in all cases, as expected due to the inclusion of zones with different thermal intensities. Notably, for short $L/D$, the growth of $\sigma_{\overline{\Nu}}$ is more pronounced, confirming that the high-$\overline{\Nu}$ core contrasts more sharply with its surroundings. At larger spacings, the increase is more gradual, indicating a smoother jet footprint, which a trend consistent with the spatial maps in~\autoref{fig:F1_Numaps}.

To further examine the spatial characteristics of the unsteady heat transfer, \autoref{fig:F3_Nurms} presents the spatial averages of the fluctuating Nusselt number $\langle \Nu' \rangle$ as a function of the normalised impingement area size $\beta$. Compared to the mean field trends, these distributions reveal distinct behaviour. For short nozzle-to-plate distances ($L/D \lesssim 2$), $\langle \Nu' \rangle$ remains approximately constant or even increases slightly in the low-$\beta$ range ($\beta \lesssim 1-2$) before gradually declining. This suggests that unsteady convective interactions in these configurations are more laterally extended across the surface. However, as $L/D$ increases, this flat region progressively diminishes, and the $\langle \Nu' \rangle$ profile begins to decay more uniformly with $\beta$, indicating that the fluctuating heat transfer becomes increasingly localised near the jet centreline. As in the mean field case, higher Reynolds numbers systematically yield higher $\langle \Nu' \rangle$ values. At the highest Reynolds tested, a mild saturation or attenuation is observed, possibly due to the reduced temporal resolution of the infrared acquisition system, as discussed previously in relation to the Fourier number.

The corresponding standard deviation, $\sigma_{\Nu'}$, also increases steadily with $\beta$, revealing growing heterogeneity in fluctuation intensity across the domain. The sensitivity to $\Ren$ is particularly noticeable at low $L/D$, where sharper fluctuation peaks result in greater spatial contrast. At large $L/D$, the increase in $\sigma_{\Nu'}$ is again more gradual, aligning with the weaker and more diffuse oscillatory patterns identified in~\autoref{fig:F1_Numaps}.

\begin{figure*}
    \centering
    \includegraphics[width=0.99\linewidth]{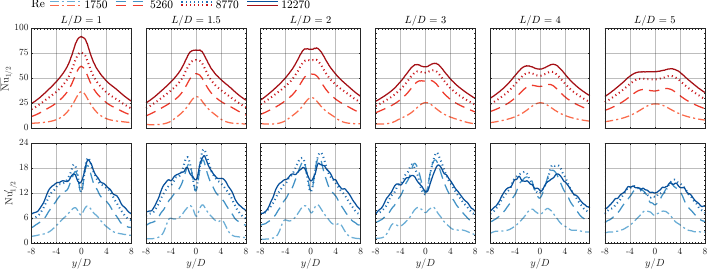}
    \caption{Spanwise profiles of the time-averaged (top row) and fluctuating (bottom row) Nusselt number at the symmetry plane $x = 0$ for all tested configurations. Each column corresponds to a different nozzle-to-plate distance $L/D$, and the colour intensity increases with Reynolds number. These profiles provide a cross-sectional view of the jet's thermal footprint and oscillatory behaviour.}
    \label{fig:F4_Numid}
\end{figure*}

% Centerline average profiles
To complement the two-dimensional maps presented in \autoref{fig:F1_Numaps} and the integral profiles in \autoref{fig:F2_Num} and \autoref{fig:F3_Nurms}, \autoref{fig:F4_Numid} offers a more focused view of the heat transfer distribution along the spanwise direction by plotting the Nusselt number profiles at the symmetry plane $x = 0$. This cross-section corresponds to the plane along which the sweeping motion occurs and is therefore particularly informative in assessing the lateral footprint and oscillatory behaviour of the jet. The top row displays the time-averaged Nusselt number profiles $\overline{\Nu}_{1/2}$, while the bottom row shows the corresponding fluctuating component $\Nu'_{1/2}$. As before, each column corresponds to a fixed $L/D$ ratio, and the colour intensity reflects the Reynolds number, with darker tones indicating higher $\Ren$.

Several trends previously identified in the full-field maps of \autoref{fig:F1_Numaps} are clearly reinforced in these spanwise profiles. For short nozzle-to-plate distances ($L/D = 1 - 1.5$), the $\overline{\Nu}$ profiles exhibit a sharp, centrally concentrated peak, indicating a highly localised stagnation region where the jet impinges in a time-averaged sense—analogous to the stagnation point in classical steady impinging jets. As $L/D$ increases, this central peak becomes flatter and broader, consistent with the progressively wider and more uniform impingement region observed in the two-dimensional maps.

The fluctuating Nusselt profiles $\Nu'_{1/2}$ reveal an even richer structure. At low to intermediate $L/D$ values ($1 \leq L/D \leq 3$), the profiles evolve into a characteristic double-lobe shape with two distinct off-centre peaks. This pattern reflects the alternating lateral impingement typical of the sweeping jet motion. Both the intensity and the spanwise separation of these peaks increase with Reynolds number, supporting the hypothesis that the sweep amplitude grows with jet momentum and oscillation frequency \citep{kim2019measurement, wen2018lagrangian}. At the highest $L/D$ values, the fluctuation intensity diminishes and the profiles become smoother, suggesting a loss of coherence in the jet’s oscillatory impact due to increased dispersion. This behaviour is consistent with the trends in \autoref{fig:F3_Nurms}, where $\langle \Nu' \rangle$ tends to plateau or slightly decline at high $\Ren$ and large $L/D$.

The mid-plane profiles compare favourably with results from the literature. \citet{zhou2019heat} reported similar spanwise distributions using a fluidic oscillator of analogous geometry but different dimensional scales. Despite minor discrepancies in absolute Nusselt values, likely attributable to geometric variations or experimental uncertainty due to the experimental setup deviations, the overall trends are robust. These findings have also been corroborated by \citet{park2018flow} and \citet{kim2019measurement}, supporting the repeatability and generality of the observed thermal behaviour.

\begin{figure*}
    \centering
    \includegraphics[width=0.99\linewidth]{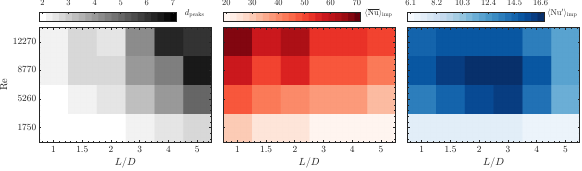}
    \caption{Condensed matrix representation of heat transfer metrics within the impingement region as a function of Reynolds number and nozzle-to-plate distance. The left panel shows the distance between the two main peaks of the fluctuating Nusselt number field ($d_{\text{peaks}}$), extracted from the $\Nu'$ distribution in the impingement zone. The middle panel displays the integrated value of the time-averaged Nusselt number ($\langle \overline{\Nu} \rangle_{\text{imp}}$) over the region where $\overline{\Nu}$ exceeds 50\% of its dynamic range. The right panel reports the integrated value of the fluctuating Nusselt number ($\langle \Nu' \rangle_{\text{imp}}$), evaluated over the area where $\Nu'$ is above 50\% of its respective dynamic range. Together, these maps provide a compact overview of the spatial structure, magnitude, and unsteadiness of convective heat transfer induced by sweeping jet impingement.}
    \label{fig:impact_matrices}
\end{figure*}

To consolidate the insights extracted from line profiles and full-field maps, \autoref{fig:impact_matrices} presents a condensed matrix visualisation of three key metrics that characterise the heat transfer behaviour across the investigated parametric space. The lateral separation between the peaks of $\Nu'$ (left panel) reveals a monotonic increase with both Reynolds number and $L/D$, confirming that higher jet momentum and longer travel distance enable broader jet deflection and footprint spreading. Conversely, the integral of the time-averaged Nusselt number within the impingement zone (middle panel) exhibits an inverse trend with $L/D$: increasing with $\Ren$ but peaking at low $L/D$. This reflects a concentration of convective activity near the stagnation region when the target surface is placed closer to the nozzle, where the jet retains higher axial coherence. Interestingly, the integral of $\Nu'$ (right panel) suggests that thermal unsteadiness is most pronounced at intermediate $L/D$ values (2--3) and high Reynolds numbers. This indicates a regime where the sweeping motion has fully developed but has not yet become overly diffused, promoting vigorous lateral motion and enhanced local mixing. At the highest $L/D$, a saturation or even decay of $\langle \Nu' \rangle_{\text{imp}}$ is observed, likely due to excessive spreading and reduced interaction intensity with the surface. Notably, the lowest Reynolds number case ($\Ren = 1750$) shows distinctly low values in both $\langle \overline{\Nu} \rangle_{\text{imp}}$ and $\langle \Nu' \rangle_{\text{imp}}$, with minimal sensitivity to $L/D$. This behaviour is consistent with the transitional nature of the flow, where the jet exhibits limited turbulent development and weaker lateral oscillation, resulting in subdued convective transport and negligible heat transfer fluctuations. These findings confirm the competing effects of sweeping coherence and turbulent dispersion in shaping the spatial and temporal characteristics of surface heat transfer, highlighting optimal parameter ranges for maximising unsteady thermal forcing.

%----- Proximity maps -----
\subsection{Data-Driven Parametric Mapping of Heat Transfer Distributions}
To conclude the analysis of the heat transfer behaviour induced by sweeping jets, we explore a low-dimensional representation of the heat transfer distributions, both time-averaged $\overline{\Nu}$ and fluctuating $\Nu'$, to gain insight into the structure of the parametric landscape and the feasibility of building surrogate models. The distributions shown in \autoref{fig:F1_Numaps} are embedded into a reduced-order space using classical Multi-Dimensional Scaling (MDS)~\citep{cox2000mds,Kaiser2017ifac}. This method projects the high-dimensional data into a two-dimensional space spanned by the coordinates $(\gamma_1, \gamma_2)$, which represent the directions of maximum dispersion based on pairwise Euclidean distances between cases. The result is a proximity map, where each marker corresponds to a given experimental condition (i.e., a unique combination of $\Ren$ and $L/D$).
\begin{figure}
    \centering
    \includegraphics[width=0.99\linewidth]{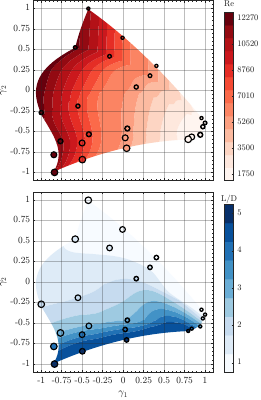}
    \caption{Low-dimensional proximity map of the spatial Nusselt number distributions based on time-averaged and fluctuating components from \autoref{fig:F1_Numaps}. The embedding is computed via classical MDS, where each marker represents a unique experimental condition. In the top figure, markers are coloured by Reynolds number, showing a dominant gradient from right to left. In the bottom figure, the same points are coloured by nozzle-to-plate distance, revealing a vertical gradient. Interpolated surfaces fitted over the convex hull approximate the parametric response manifold, enabling interpolated estimation of heat transfer fields from the reduced space.}
    \label{fig:F21_proximity}
\end{figure}

The resulting projection, displayed in \autoref{fig:F21_proximity}, reveals a clear and interpretable organisation of the data. In the top panel, the colour scale indicates the Reynolds number. A dominant gradient from right to left is observed, capturing the transition from low to high momentum cases. Interestingly, configurations with low $\Ren$ collapse into a narrow region, reflecting the relatively minor differences in heat transfer distributions at low flow rates, which is consistent with the similarity observed in the first row of \autoref{fig:F1_Numaps}. As $\Ren$ increases, the embedded points diverge, highlighting the growing influence of nozzle-to-plate distance.
This effect is emphasised in the bottom panel of \autoref{fig:F21_proximity}, where the same embedded space is coloured according to $L/D$. Here, a distinct vertical ordering of the points is observed, particularly at high Reynolds numbers, indicating that the proximity map effectively decouples the two control parameters. While low $\Ren$ cases remain clustered, at higher $\Ren$ the points progressively shift along the $\gamma_2$ axis with increasing $L/D$, reflecting the more significant impact of nozzle-to-plate distance on the heat transfer distributions under these conditions.

To complement the interpretability of the MDS embedding and explore its use as a predictive tool, a surrogate modelling framework is developed capable of reconstructing the full heat transfer fields $(\overline{\Nu}, \Nu')$ based solely on the parametric inputs $(\Ren, L/D)$. This framework is based on a two-step data-driven architecture: a forward projection from physical parameters to the embedded manifold coordinates using a Thin Plate Spline (TPS) regressor, followed by a local back-mapping from the embedded space to the high-dimensional thermal field via first-order approximation.

The forward regressor operates in the two-dimensional embedded space, learning a mapping function $f_r: (\Ren, L/D) \in \mathbb{R}^2 \rightarrow (\gamma_1,\gamma_2) \in \mathbb{R}^2$. This regression task is significantly simpler than learning a direct mapping from $(\Ren, L/D)$ to high-dimensional Nusselt distributions. For this purpose, a Radial Basis Function (RBF) interpolator is employed using a thin plate spline kernel of the form $r^2 \log(r)$, which is particularly well-suited for interpolating scattered planar data. The implementation is based on the \texttt{RBFInterpolator} method from the \texttt{SciPy} Python library.

While the MDS algorithm preserves pairwise distances between cases in the reduced space, it does not provide a direct inverse mapping. To reconstruct the full thermal field from a point $(\gamma_1,\gamma_2)$ in the manifold, a local back-mapping is used based on a $\kappa$-nearest neighbours ($\kappa$-NN) approach, following the method described in~\citet{Farzamnik_2023_jfm}. Let $\mathbf{x}^{(i)}:= (\overline{\Nu}, \Nu') \in \mathbb{R}^P$ denote a high-dimensional data point, where $P = 2 \times n_x \times n_y$ is the number of spatial degrees of freedom. Its corresponding low-dimensional embedding is $\mathbf{y}^{(i)} := (\gamma_1, \gamma_2) \in \mathbb{R}^2$. The goal is to estimate the back-mapping $f_b: \mathbb{R}^2 \rightarrow \mathbb{R}^P$ for any $\mathbf{y}$ in the embedded domain.
The reconstruction $\mathbf{x}$ associated with a new point $\mathbf{y}$ is obtained via a first-order Taylor expansion about its nearest neighbour $\mathbf{y}^{(1)}$, using
\begin{equation}
    \mathbf{x} = \mathbf{x}^{(1)} + \nabla f\big(\mathbf{y}^{(1)} \big) \left(\mathbf{y} - \mathbf{y}^{(1)}\right), 
\end{equation}
where the Jacobian matrix $\nabla f(\mathbf{y}^{(1)})$ is estimated from the remaining $\kappa-1$ neighbours via a least-squares fit:
\begin{equation}\label{eq:knnproj}
    \begin{bmatrix}
        \mathbf{x}^{(2)} -  \mathbf{x}^{(1)}\\
        \ldots \\
        \mathbf{x}^{(\kappa)} -  \mathbf{x}^{(1)}
        \end{bmatrix}
        \simeq
        \nabla f(\mathbf{y}^{(1)})
        \begin{bmatrix}
        \mathbf{y}^{(2)} -  \mathbf{y}^{(1)}\\
        \ldots \\
        \mathbf{y}^{(\kappa)} -  \mathbf{y}^{(1)}
    \end{bmatrix},%
\end{equation}
leading to the closed-form least-squares solution
\begin{equation}
    \nabla f(\mathbf{y}^{(1)}) = \left( \Delta \mathbf{Y}^\top \Delta \mathbf{Y} \right)^{-1} \Delta \mathbf{Y}^\top \Delta \mathbf{X},
\end{equation}
with $\Delta \mathbf{X}$ and $\Delta \mathbf{Y}$ being the centered neighbour matrices defined from Eq.~\eqref{eq:knnproj}.

To ensure generalisation without extrapolation, the surrogate was validated using carefully selected train-test splits. The training set always includes the boundary cases (i.e., combinations with extreme values of $\Ren$ and $L/D$) to avoid extrapolating outside the convex hull in the parametric space. From the remaining interior conditions, 8 representative splits were tested, each using 2 test cases and 22 training cases. In all splits, the MDS embedding constructed from the training set alone closely reproduced the global manifold structure observed in the full dataset (\autoref{fig:F21_proximity}), confirming the convergence of the reduced representation based on the current dataset. Ultimately, the back-mapping step was performed using $\kappa = 4$ nearest neighbours, which yielded the best trade-off between local accuracy and generalisation across the tested splits.

The predictive performance of the surrogate model was evaluated over eight train/test splits, as summarised in Table~\ref{tab:surrogate_metrics}. Metrics are reported separately for the time-averaged ($\overline{\Nu}$) and fluctuating ($\Nu'$) Nusselt number distributions, although both fields are predicted simultaneously from a shared MDS embedding. Across all splits, the mean absolute error (MAE) remains below 1.6 for $\overline{\Nu}$ and 0.55 for $\Nu'$, with corresponding root mean squared errors (RMSE) under 2.0 and 0.7, respectively. These values indicate that the surrogate model captures both large-scale spatial features and localised variations with high accuracy. The coefficient of determination ($R^2$) consistently exceeds 0.95 for both components, implying that more than 95\% of the variance in the reference distributions is retained in the predictions. Pearson correlation coefficients above 0.99 for $\overline{\Nu}$ and 0.98 for $\Nu'$ confirm excellent agreement in the overall spatial structure. Notably, the Structural Similarity Index Measure (SSIM), which evaluates perceptual and textural similarity in two-dimensional fields, achieves average scores of 0.935 and 0.887 for $\overline{\Nu}$ and $\Nu'$, respectively. These results confirm that the reconstructed fields preserve not only amplitude accuracy but also spatial coherence and morphology relevant to physical interpretation. Overall, the surrogate model demonstrates robust predictive capability across a range of flow conditions, with low dispersion across splits (see standard deviations in Table~\ref{tab:surrogate_metrics}). This validates the proposed MDS–TPS–backmapping pipeline as a viable strategy for predicting the joint $(\overline{\Nu}, \Nu')$ fields using only the flow parameters $(\Ren, L/D)$.

%- Table with Metrics
\begin{table}[]
\caption{Performance metrics of the surrogate model evaluated over eight train/test splits. Each row corresponds to the average performance over test samples in a given split, while the last two rows report the mean and standard deviation across all splits. The top table summarises the prediction of time-averaged Nusselt number fields ($\overline{\Nu}$), while the bottom one refers to the fluctuating component ($\Nu'$). Metrics include the mean absolute error (MAE), root mean squared error (RMSE), mean squared error (MSE), coefficient of determination ($R^2$), Pearson correlation coefficient (Corr), and structural similarity index measure (SSIM).}
\label{tab:surrogate_metrics}
% Num
\begin{tabular}{cccccc}
\hline
\textbf{Model} & \textbf{MAE} & \textbf{RMSE} & \textbf{$R^2$} & \textbf{Corr} & \textbf{SSIM} \\ \hline
1 & 1.170 & 1.517 & 0.974 & 0.992 & 0.900 \\
2 & 0.920 & 1.188 & 0.983 & 0.995 & 0.894 \\
3 & 1.337 & 1.700 & 0.971 & 0.995 & 0.893 \\
4 & 1.088 & 1.336 & 0.978 & 0.993 & 0.946 \\
5 & 1.555 & 1.948 & 0.963 & 0.993 & 0.929 \\
6 & 1.318 & 1.619 & 0.959 & 0.988 & 0.948 \\
7 & 1.283 & 1.547 & 0.971 & 0.995 & 0.977 \\
8 & 1.002 & 1.269 & 0.978 & 0.997 & 0.990 \\ \hline
\textbf{Mean}  & \textbf{1.209}  & \textbf{1.515}  & \textbf{0.972}   & \textbf{0.993}   & \textbf{0.935}  \\
\textbf{STD}   & \textbf{0.206}  & \textbf{0.249}  & \textbf{0.008}  & \textbf{0.003}  & \textbf{0.037}  \\ \hline
\end{tabular}

\vspace{0.3cm}
%Nuf
\begin{tabular}{ccccccc}
\hline
\textbf{Model} & \textbf{MAE} & \textbf{RMSE} & \textbf{$R^2$} & \textbf{Corr} & \textbf{SSIM} \\ \hline
1 & 0.405 & 0.539 & 0.977 & 0.990 & 0.860 \\
2 & 0.382 & 0.557 & 0.973 & 0.991 & 0.884 \\
3 & 0.529 & 0.695 & 0.956 & 0.989 & 0.861 \\
4 & 0.322 & 0.448 & 0.982 & 0.994 & 0.892 \\
5 & 0.541 & 0.680 & 0.957 & 0.989 & 0.824 \\
6 & 0.363 & 0.496 & 0.976 & 0.993 & 0.924 \\
7 & 0.448 & 0.577 & 0.964 & 0.993 & 0.922 \\
8 & 0.490 & 0.661 & 0.952 & 0.994 & 0.931 \\ \hline
\textbf{Mean} & \textbf{0.435} & \textbf{0.582} & \textbf{0.967} & \textbf{0.992} & \textbf{0.887} \\
\textbf{STD}  & \textbf{0.080} & \textbf{0.090} & \textbf{0.011} & \textbf{0.002} & \textbf{0.038} \\ \hline
\end{tabular}
\end{table}

\begin{figure*}
    \centering
    \includegraphics[width=0.95\linewidth]{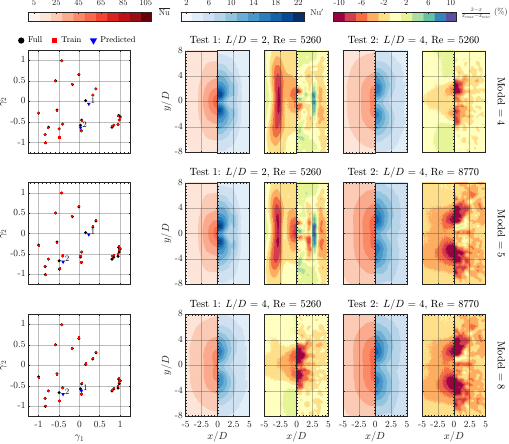}
    \caption{Qualitative assessment of surrogate model performance for three representative models (rows), each tested on two unseen cases. The leftmost column shows the MDS embedding: full dataset (black circles), training points (red squares), and surrogate-predicted locations for the test cases (blue triangles). For each test case, the left subpanel displays the predicted time-averaged ($\overline{\Nu}$, left half) and fluctuating ($\Nu'$, right half) heat transfer fields. The right subpanel shows the spatially resolved relative error (\%) of each predicted field with respect to its corresponding ground truth, normalised by the full range of the reference field. This allows direct comparison between fields with different magnitudes. Test cases are chosen to span different regions of the parameter space and to evaluate robustness across varying flow regimes. Colour scales match those of \autoref{fig:F1_Numaps}.}
    \label{fig:Surrogate_maps}
\end{figure*}

To complement the global metrics reported in Table~\ref{tab:surrogate_metrics}, \autoref{fig:Surrogate_maps} presents a visual comparison between surrogate-predicted and reference fields for three representative models. Each row corresponds to a different train/test split, with two test cases per model selected to span contrasting regions of the parametric space. The proximity maps on the left confirm that the manifold geometry obtained from the training data closely matches the full-data embedding, even when excluding adjacent interior points. The blue triangles marking the predicted test coordinates lie on smooth interpolations of the training manifold, illustrating the effectiveness of the Thin Plate Spline regression. The subpanels show the predicted distributions of time-averaged and fluctuating Nusselt numbers alongside their spatially resolved relative errors. The first test case ($\Ren = 5260$, $L/D = 2$) highlights the surrogate’s strengths and limitations in the transitional regime. The time-averaged Nusselt number is slightly underestimated in the jet impact region, with relative errors up to $-8\%$, while the surrounding footprint reaches $-10\%$. Fluctuating heat transfer ($\Nu'$) is reconstructed with good spatial agreement, although minor discrepancies arise at the interface between the two characteristic lobes of the sweeping jet and the external region. These patterns are consistent across Models 4 and 5, confirming robustness against train/test selection. The second test case in Models 5 and 8 ($\Ren = 8770$, $L/D = 4$) represents a more complex scenario with increased turbulence intensity and lateral jet deflection. Here, both time-averaged and fluctuating fields are consistently underestimated by the surrogate. The largest errors occur near the extrema of the sweeping lobes—regions associated with jet reversal, suggesting that the surrogate struggles to reconstruct high-gradient features in these dynamically sensitive zones fully. Moreover, elevated local errors in $\Nu'$ are exacerbated by the increased noise level at high $\Ren$, a known limitation due to reduced signal-to-noise ratio in time-resolved IR measurements. Finally, the remaining test case ($\Ren = 5260$, $L/D = 4$) demonstrates a stronger predictive performance. The time-averaged distribution is well approximated, with relative errors confined to the range $-5\%$ to $+3\%$, indicating accurate estimation of both peak magnitude and spatial footprint. For the fluctuating field, underestimation is observed primarily within the central impingement region, with residual noise affecting the peripheral areas. 

Concluding, in all cases, the surrogate reconstructs both global structure and finer spatial gradients with high fidelity. The largest discrepancies occur in peripheral regions of the impingement zone, particularly for the fluctuating component, yet these remain within $\pm10\%$ in relative terms. These visual impressions are consistent with the high SSIM values and low MAE and RMSE metrics reported in Table~\ref{tab:surrogate_metrics}, reinforcing the model's ability to generalise across unseen Reynolds numbers and nozzle-to-plate distances. This confirms the feasibility of the proposed low-dimensional, data-driven framework for rapid surrogate prediction of heat transfer behaviour.

%----- PIV results -----
\subsection{Flow Field Measurements}
\begin{figure*}
    \centering
    \includegraphics[width=0.99\linewidth]{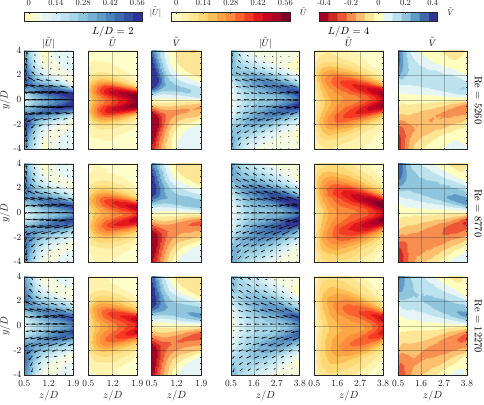}
    \caption{Time-averaged velocity fields for two nozzle-to-plate distances ($L/D = 2$ and $4$) and three Reynolds numbers ($\Ren = 5260$, $8770$, and $12270$). Each row corresponds to a fixed Reynolds number and each column to a specific quantity: velocity magnitude $|\tilde{\mathbf{U}}|$ (left), streamwise component $\tilde{U}$ (middle), and vertical component $\tilde{V}$ (right). Velocity components are normalised by the characteristic velocity $U_c$ derived from \autoref{fig:F1_Freq}. Vector quivers are overlaid on the $|\tilde{\mathbf{U}}|$ fields to aid interpretation. The wall is located at $z=0$.}
    \label{fig:F11_PIVavg_maps}
\end{figure*}

To complement the heat transfer measurements and better understand the underlying flow mechanisms, we now turn to the time-averaged velocity fields obtained from PIV measurements. Figure~\ref{fig:F11_PIVavg_maps} presents the non-dimensional velocity components $\tilde{U}$, $\tilde{V}$, and the magnitude $|\tilde{\mathbf{U}}|$, each normalised with the characteristic velocity $U_c$ defined in Figure~\ref{fig:F1_Freq}. The analysis focuses on two intermediate nozzle-to-plate distances ($L/D = 2$ and $L/D = 4$), representative of the transitional regime between highly confined and fully dispersed jet behaviours. For each $L/D$, three Reynolds numbers, excluding the lowest case from previous analysis to ensure adequate seeding density and flow resolution within the PIV field of view.

The $\tilde{U}$ fields display a canonical sweeping jet structure across all cases: a high-speed core emerges from the nozzle exit, reaching peak velocities of approximately $0.6\,U_c$, consistent with the expected ratio between throat and outlet areas. As the jet convects downstream, this high-speed core bifurcates symmetrically, forming two lobes associated with the two extreme positions of the sweeping motion. The $\tilde{U}$ component diminishes along $z/L_c$, approaching zero near the wall due to the impingement and redirection of the jet. This velocity decay near the surface is physically consistent with the lower values of $\overline{\Nu}$ observed in the peripheral regions of the heat transfer maps (see \autoref{fig:F1_Numaps}), where the jet loses momentum and thus its capacity to sustain strong convective cooling.

The $\tilde{V}$ component provides further insight into the sweeping dynamics. As expected from symmetry and temporal averaging, $\tilde{V}$ is nearly zero along the central $y/L_c = 0$ axis. However, off-center, two lateral zones of opposite vertical velocity appear, corresponding to the alternating impact of the jet during its sweeping motion. Notably, these zones exhibit increasing intensity closer to the impingement surface, indicating that the jet’s streamwise momentum is redirected into lateral motion upon impact. This redirection is more abrupt at lower $L/D$, leading to sharper vertical gradients near the wall. In contrast, at $L/D = 4$, the momentum redistribution is smoother and more spatially diffused. These flow features are consistent with the spanwise heat transfer profiles (\autoref{fig:F3_Nurms}), where low $L/D$ produces sharper $\Nu'$ peaks due to more focused lateral sweeping, whereas high $L/D$ cases exhibit broader, less distinct fluctuation patterns.

The maps of $|\tilde{\mathbf{U}}|$, enriched with vector quivers, synthesise the behaviour of both components. The quivers illustrate the flow’s deflection and spanwise spreading as it approaches the plate. At lower $L/D$, this redirection occurs within a shorter axial distance, resulting in more intense surface interaction and a more confined thermal footprint. At higher $L/D$, the longer travel allows for further jet dispersion before impact, consistent with the broader $\overline{\Nu}$ and attenuated $\Nu'$ profiles reported earlier.

\begin{figure*}
    \centering
    \includegraphics[width=0.99\linewidth]{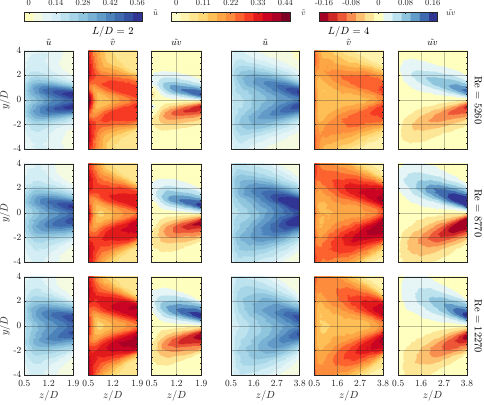}
    \caption{Fluctuating velocity fields (rms values) for two nozzle-to-plate distances ($L/D = 2$ and $4$) and three Reynolds numbers ($\Ren = 5260$, $8770$, and $12270$). Each row corresponds to a fixed Reynolds number. The columns show: (left) streamwise velocity fluctuations $\tilde{u}$, (middle) vertical velocity fluctuations $\tilde{v}$, and (right) Reynolds shear stress $\widetilde{uv}$. All quantities are normalised by the characteristic velocity $U_c$. The wall is located at $z=0$.}
    \label{fig:F12_PIVfluc_maps}
\end{figure*}

Overall, the time-averaged velocity fields reinforce the observed heat transfer trends and provide a physical explanation for the spatial features seen in the Nusselt maps. The redistribution of momentum near the surface, governed by both $L/D$ and $\Ren$, directly influences the magnitude and localisation of the convective heat transfer response. However, understanding the full behaviour of sweeping jets requires not only the mean flow structure but also its unsteady dynamics, which are central to the generation of heat transfer fluctuations.
To this end, \autoref{fig:F12_PIVfluc_maps} presents the root-mean-square (rms) distributions of the normalised fluctuating velocity components $\tilde{u}$ and $\tilde{v}$, along with the Reynolds shear stress $\widetilde{uv}$, for the same cases as in \autoref{fig:F11_PIVavg_maps}. These fields complement the mean flow analysis by capturing the energetic oscillations that arise from the jet's periodic sweeping motion, and they provide a direct connection to the fluctuation-dominated heat transfer patterns discussed in Section~\ref{ss:heattransfer}.

The streamwise fluctuations $\tilde{u}$ exhibit two pronounced off-center peaks for all tested Reynolds numbers, with the strongest amplitudes observed at $L/D = 2$. These peaks align with the outer sweep positions of the jet, where the momentum reversal is most abrupt. As the jet oscillates between these lateral extremes, strong streamwise accelerations and decelerations occur, especially in the vicinity of the stagnation line, resulting in elevated $\tilde{u}$ values. At $L/D = 4$, these peaks remain discernible but become broader and less intense, consistent with the smoother, more gradual redirection of the jet as it has more space to spread before impingement. These lateral peaks of $\tilde{u}$ correlate well with the locations of $\Nu'$ maxima in the IR-based maps (cf. \autoref{fig:F1_Numaps}), confirming that the primary source of thermal fluctuations originates from the unsteady jet impingement zones.

The wall-tangent fluctuations $\tilde{v}$ also reveal interesting trends. While less localised than $\tilde{u}$, their magnitude increases with Reynolds number and shows broader lateral coverage at $L/D = 4$. This suggests that the vertical component of unsteadiness contributes to a wider influence area on the wall, consistent with the larger span of fluctuating Nusselt number observed in the heat transfer maps. Moreover, the spatial symmetry and outward shift of $\tilde{v}$ with increasing $\Ren$ indicates a stronger sweeping amplitude and more energetic vortex dynamics. These observations are in line with the findings of \citet{dangelo2024piv} who reported similar trends in jet spreading and unsteadiness with increasing $L/D$. In their detailed PIV study, the fluctuation levels of $v$ were shown to extend deeper into the wall-normal direction and across a wider lateral range as $L/D$ increased, confirming that the herein IR-based heat transfer observations are supported by independent velocity field assessments.

The Reynolds shear stress distributions, $\widetilde{uv}$, further support the interpretation of shear-driven unsteadiness in the jet’s near-field. Across all cases, two lobes emerge symmetrically on either side of the jet centreline. These regions indicate intense shear and momentum exchange across the spanwise direction and coincide spatially with the regions of maximum $\tilde{u}$. The $\widetilde{uv}$ magnitude peaks are more compact and stronger for $L/D = 2$, while they become broader and shift downstream at $L/D = 4$. This spatial redistribution mirrors the shift in $\Nu'$ intensity observed earlier and suggests a spanwise migration of the coherent structures responsible for lateral momentum transport. These results are consistent with the analysis presented by \citet{dangelo2024piv}, who observed a downstream displacement and weakening of shear-layer instabilities with increasing $L/D$, especially at intermediate Reynolds numbers.

Altogether, the spatial organisation and evolution of the fluctuating velocity components provide a compelling mechanistic explanation for the distribution of $\Nu'$ and its dependence on both Reynolds number and nozzle-to-plate spacing. High $\tilde{u}$ and $\widetilde{uv}$ levels are directly linked to localised thermal fluctuation hotspots, while broader $\tilde{v}$ fields explain the expanded, though weaker, heat transfer footprints at large $L/D$. The combined PIV-IR approach used here confirms and complements the results in the literature \citep{dangelo2024piv} while extending the interpretation to include thermal implications \citep{dangelo2025ir, PIRT_2025}, thus offering a more complete understanding of the physics governing impinging sweeping jets.

%----- Comparison of IR and PIV data ------
\subsection{Relating Velocity and Heat Transfer Fluctuations}
\begin{figure*}
    \centering
    \includegraphics[width=0.85\linewidth]{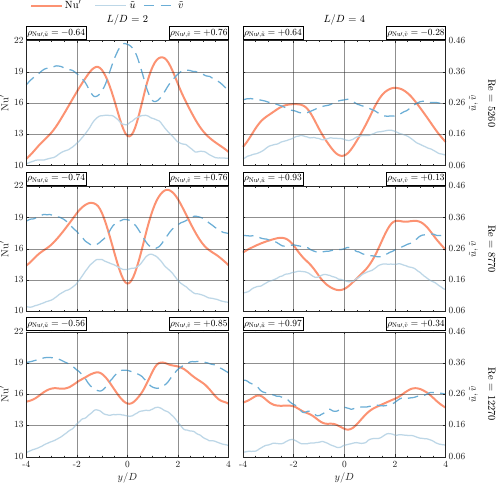}
    \caption{Comparison between fluctuating Nusselt number ($\Nu'$) and velocity fluctuations at the wall along the symmetry plane $x = 0$ for different Reynolds numbers (rows) and nozzle-to-plate distances $L/D$ (columns). Profiles of $\tilde{u}$ (solid blue) and $\tilde{v}$ (dashed blue) are extracted from the wall-parallel intersection of the PIV measurement plane, while $\Nu'$ (solid red) is obtained from the IR measurements. The Pearson correlation coefficients between $\Nu'$ and each fluctuating velocity component are reported.}
    \label{fig:F13_midprof_fluc}
\end{figure*}

To directly relate the velocity field dynamics to the surface heat transfer behaviour, we compare in~\autoref{fig:F13_midprof_fluc} the spatial profiles of the fluctuating Nusselt number $\Nu'$ and the velocity fluctuations $\tilde{u}$ and $\tilde{v}$ at the intersection line between the IR and PIV measurement planes, i.e. the symmetry plane $x=0$ at the target wall ($z=0$). This profile-based analysis offers a localised and spatially-correlated interpretation of the flow–thermal coupling in the sweeping jet system, highlighting how velocity fluctuations govern the convective heat transfer at the wall.
Across all cases, $\Nu'$ exhibits a pronounced double-peak structure, most visible at $L/D = 2$, where the sweeping motion impinges more directly onto the wall. This double-lobed distribution is mirrored by both velocity fluctuation components, particularly $\tilde{v}$, which shows strong symmetry and coherence at lower $L/D$. These results highlight the dominant role of wall-normal oscillations in periodically disturbing the thermal boundary layer, promoting intense localised heat transfer fluctuations.

As the nozzle-to-plate spacing increases to $L/D = 4$, the $\Nu'$ profiles flatten and broaden, reflecting a more distributed and less intense oscillatory heat transfer pattern. Correspondingly, the velocity fluctuations change in character: while $\tilde{v}$ becomes more spatially uniform and less structured, $\tilde{u}$ grows in relative prominence, especially at higher Reynolds numbers. This shift suggests that for larger $L/D$, lateral sweep and shear-driven unsteadiness in the streamwise direction become the primary contributors to heat transfer modulation, rather than the direct wall-normal jet impact.

The qualitative agreement between $\Nu'$ and velocity fluctuation profiles is reinforced by the correlation trends reported in each subplot. At low $L/D$, the stronger correspondence with $\tilde{u}$ confirms that convective fluctuations are driven by periodic wall-normal injection. In contrast, for large $L/D$, the modulation of heat transfer is better aligned with spanwise unsteadiness, captured by $\tilde{v}$, particularly under high $\Ren$ conditions where sweep amplitude and inertia are greatest.

These observations reflect the underlying flow structure captured in Figures~\ref{fig:F11_PIVavg_maps} and~\ref{fig:F12_PIVfluc_maps}, and explain the change in shape and magnitude of $\Nu'$ observed across $L/D$ and $\Ren$. This cross-analysis confirms that both $\tilde{u}$ and $\tilde{v}$ are relevant contributors to convective fluctuation, but their relative importance depends strongly on the impingement geometry. Furthermore, the existence of a high degree of correlation for specific configurations reinforces the potential for velocity-based estimators of thermal fluctuations, laying a promising foundation for reduced-order modelling or control-oriented strategies in sweeping jet systems.

%%%%%%%%%%%%%%%%%%%%%%%%%%%%%%%%%%%%%%%%%%%%%%%%%%%%%%%%%%%%%%%
%%%%%%%%%%%%%%%%%%%%%%%%%%% CONCLUSIONS %%%%%%%%%%%%%%%%%%%%%%%
%%%%%%%%%%%%%%%%%%%%%%%%%%%%%%%%%%%%%%%%%%%%%%%%%%%%%%%%%%%%%%%
\section{Conclusions \label{s:Conclusions}}
This study has presented a comprehensive experimental investigation of the convective heat transfer and flow dynamics of an impinging sweeping jet, with a particular focus on the effects of Reynolds number and nozzle-to-plate distance. By combining non-synchronous time-resolved infrared thermography with planar particle image velocimetry, a detailed and spatially correlated characterisation of both the mean and fluctuating behaviour of the jet has been achieved.

The time-averaged Nusselt number distribution on the impinging surface reveal the well-known influence of Reynolds number on convective performance and show that the jet's thermal footprint broadens as the nozzle-to-plate distance increases, leading to more uniform but less intense heat transfer. More importantly, the fluctuating Nusselt fields unveil rich unsteady structures that cannot be inferred from average data alone. These structures reflect the oscillatory nature of the sweeping jet and localise the thermal unsteadiness at the lateral sweep extremes. Despite being often neglected in prior studies, the unsteady contribution $\Nu'$ is shown to account for up to 10–15\% of the mean value, underscoring its non-negligible role in surface cooling and its importance in the thin-foil energy balance.

The parametric study quantifies the effect of Reynolds number and $L/D$ on the impingement region, summarised by three key figures of merit: the distance between the lobes (peaks of $\Nu'$), the integrated time-averaged Nusselt number within the impingement zone, and the integral of fluctuating Nusselt number. The distance between the lobes consistently increases with Reynolds number, reflecting the broader lateral jet deflection at higher flow rates. For time-averaged Nusselt number, larger $Re$ values result in higher peak heat transfer, but this benefit diminishes as $L/D$ increases, with the heat transfer becoming more uniform across the surface. Conversely, fluctuating Nusselt number distributions show that the unsteady component increases with both $Re$ and $L/D$, peaking at intermediate values.

Velocity field measurements provide the necessary physical interpretation for these thermal patterns. The time-averaged streamwise velocity confirms the presence of a bifurcated jet core, while the wall-normal velocity reveals the characteristic lateral sweep dynamics. Fluctuating velocity fields expose the oscillatory motion's intensity and structure, with streamwise and wall-normal components contributing differently depending on the nozzle-to-plate distance. The Reynolds shear stress maps further reinforce the existence of shear-driven momentum exchange that coincides spatially with the $\Nu'$ maxima.

A particularly novel aspect of this work lies in the one-to-one spatial comparison of $\Nu'$ and velocity fluctuation profiles along the symmetry plane. The observed correlations confirm that wall-normal velocity fluctuations dominate the thermal response at low $L/D$, whereas streamwise fluctuations become increasingly relevant as the jet disperses before impingement. These results not only validate the physical connection between unsteady flow and unsteady heat transfer, but also open the door for velocity-informed surrogate models of surface thermal behaviour.

In addition, by embedding the full set of heat transfer distributions into a reduced-order space via multidimensional scaling, we demonstrate that the complex parametric behaviour of the system can be represented in a low-dimensional manifold. This data-driven framework captures the combined effect of Reynolds number and nozzle-to-plate spacing and offers a practical tool for predicting and interpolating convective heat transfer response without the need for full-field measurements. The predictive accuracy of the surrogate model, based on Thin Plate Spline interpolation within the MDS space, was quantitatively assessed across eight train/test splits. The MAE for $\overline{\Nu}$ is consistently below 1.6, and for $\Nu'$, it remains under 0.55. RMSE values for both fields are also low: under 2.0 for $\overline{\Nu}$ and 0.7 for $\Nu'$. The coefficient of determination ($R^2$) exceeds 0.95, ensuring high fidelity of the surrogate predictions, and the SSIM values indicate that spatial features are well-preserved, with values above 0.93 for $\overline{\Nu}$ and 0.88 for $\Nu'$. Notably, the relative error in spatial predictions remains below $\pm10\%$ for all test cases, demonstrating robust generalisation across Reynolds numbers and nozzle-to-plate distance variations.

In conclusion, this work highlights the critical role of unsteady convective effects in sweeping jet systems and the value of coupling infrared thermography with PIV to fully characterise their dynamics. The methodology adopted here, including the use of spatio-temporally filtered IR data and data-driven embedding techniques, provides a robust framework for future studies on flow control, active cooling strategies, and reduced-order modelling of unsteady heat transfer systems. These findings confirm the feasibility of using the surrogate model for predicting heat transfer fields with minimal input parameters and show that the dominant thermal and dynamic features of sweeping jets can be captured using just two parameters: Reynolds number and nozzle-to-plate spacing. This opens the door for efficient predictive modelling and optimised design of sweeping jet systems.

%%%%%%%%%%%%%%%%%%%%%%%%%%%%%%%%%%%%%%%%%%%%%%%%%%%%%%%%%%%%%%%%%%%%%
%%%%%%%%%%%%%%%%%%%%%%%%%%% Acknowledgments %%%%%%%%%%%%%%%%%%%%%%%%%%%%%
%%%%%%%%%%%%%%%%%%%%%%%%%%%%%%%%%%%%%%%%%%%%%%%%%%%%%%%%%%%%%%%%%%%%%
\section*{Acknowledgments}
The work has been supported by the project EXCALIBUR (ref. PID2022-138314NB-I00), funded by MCIU/AEI/ 10.13039/501100011033 and by ‘ERDF A way of making Europe’. The authors thank Professor A. Ianiro for his useful comments and suggestions, and V. Duro and I. Robledo for the experimental setup, support and discussion. 

\section*{Data availability}
Data will be made available upon request. The InfraRed thermography processing toolbox is available in Github \hyperlink{https://github.com/rcastellanosgdb/PIRT}{github.com/rcastellanosgdb/PIRT}.

\section*{Declaration of generative AI and AI-assisted technologies in the writing process}
During the preparation of this work, the authors used ChatGPT (OpenAI) and Grammarly to check grammar, enhance readability, and improve text clarity. After using these tools, the authors reviewed and edited the content as needed and take full responsibility for the content of the publication.

\printcredits % el CRediT

%%%%%%% BIB
\bibliographystyle{model1-num-names}
\bibliography{main.bib}

\end{document}